\documentclass[aps,prl,superscriptaddress,showpacs,floatfix,nofootinbib,notitlepage,twocolumn]{revtex4-1}
\usepackage{amsmath,graphicx,float,csquotes,mdframed,appendix,url}
\usepackage[colorlinks=true, pdfstartview=FitV, linkcolor=red, citecolor=blue, urlcolor=blue]{hyperref}

\def\Glauber{\textsc{3d-glauber}}
\def\MUSIC{\textsc{music}}
\def\UrQMD{\textsc{urqmd}}
\def\iEBEMUSIC{\textsc{iebe-music}}
\def\GlauberMUSICUrQMD{\textsc{3d-glauber+music+urqmd}}

\begin{document}

\title{Collectivity in Ultra-Peripheral Pb+Pb Collisions at the Large Hadron Collider}

\author{Wenbin Zhao}
\affiliation{Department of Physics and Astronomy, Wayne State University, Detroit, Michigan 48201, USA}

\author{Chun Shen}
\affiliation{Department of Physics and Astronomy, Wayne State University, Detroit, Michigan 48201, USA}
\affiliation{RIKEN BNL Research Center, Brookhaven National Laboratory, Upton, NY 11973, USA}

\author{Bj\"orn Schenke}
\affiliation{Physics Department, Brookhaven National Laboratory, Upton, NY 11973, USA}

\begin{abstract}
We present the first full (3+1)D dynamical simulations of ultra-peripheral Pb+Pb collisions at the Large Hadron Collider. Extrapolating from p+Pb collisions, we explore whether a quasi-real photon $\gamma^*$ interacting with the lead nucleus in an ultra-peripheral collision can create a many-body system exhibiting fluid behavior. Assuming strong final-state interactions, we provide model results for charged hadron multiplicity, identified particle mean transverse momenta, and charged hadron anisotropic flow coefficients, and compare them with experimental data from the ALICE and ATLAS collaborations. The elliptic flow hierarchy between p+Pb and $\gamma^*$+Pb collisions is dominated by the difference in longitudinal flow decorrelations and reproduces the experimental data well. We have demonstrated that our theoretical framework provides a quantitative tool to study particle production and collectivity for all system sizes, ranging from central heavy-ion collisions to small asymmetric collision systems at the Relativistic Heavy-Ion Collider and the Large Hadron Collider and even at the future Electron-Ion Collider.
\end{abstract}

\maketitle

%%%% Introduction
\noindent {\it 1. Introduction. }
Ultra-relativistic collisions of heavy ions create and allow for the study of a novel state of matter, the Quark-Gluon Plasma (QGP), which exhibits the degrees of freedom of the fundamental building blocks of visible matter. Precise measurements of  the emergent collectivity of the produced matter were performed at the Relativistic Heavy Ion Collider (RHIC) and the Large Hadron Collider (LHC).  Based on the success of hydrodynamic descriptions, extensive phenomenological studies have revealed that the created QGP fireball behaves like a nearly perfect fluid with very small specific shear viscosity (see reviews~\cite{Heinz:2013th, Gale:2013da, Shen:2015msa}).

Rather strikingly, many features of collective expansion have also been observed in very small collision systems, such as p+Au, d+Au, $^3$He+Au at RHIC~\cite{PHENIX:2017xrm,PHENIX:2018lia}, and p+p and p+Pb collisions at the LHC~\cite{Li:2012hc,Dusling:2015gta,Nagle:2018nvi}.  The flow-like signals in the small systems can be interpreted as a result of the hydrodynamic response of the QGP medium to the initial collision geometry~\cite{ Nagle:2013lja,Schenke:2014zha,Shen:2016zpp,Weller:2017tsr,Mantysaari:2017cni,Zhao:2017rgg,Bierlich:2019wld,Zhao:2020wcd}. Alternatively, the Color Glass Condensate (CGC) effective theory has also predicted a significant amount of correlations in the initial state of these small collision systems, that can mimic collective behavior to a certain degree \cite{Dusling:2012cg,Lappi:2015vta,Schenke:2016lrs,Dusling:2017dqg,Dusling:2017aot}. Theoretical frameworks including both initial-state correlations and final-state interactions have been used to better understand the origin of the observed collective behavior in these small systems \cite{Schenke:2019pmk,Giacalone:2020byk,Schenke:2021mxx}.

Recently, intriguing experimental results on two-particle azimuthal correlations in ultra-peripheral Pb+Pb collisions (UPCs) at the LHC~\cite{ATLAS:2021jhn} have appeared. UPCs have appreciable rates of photo-nuclear interactions~\cite{Bertulani:2005ru,Baltz:2007kq}, and the ATLAS measurements of such photo-nuclear ($\gamma^*$+Pb) interactions in Pb+Pb UPCs indicate the persistence of collective phenomena with the strength of correlations comparable to that observed in proton-proton and proton-lead collisions in similar multiplicity ranges~\cite{ATLAS:2021jhn}.

Quantitative understanding of the many-body dynamics in UPC events poses big challenges to the theory community. First,  it is known that in asymmetric systems boost-invariance is strongly broken \cite{Bozek:2015swa,Ke:2016jrd,Schenke:2016ksl,Shen:2020jwv,Wu:2021hkv}. Because of the largely different incoming energies between the quasi-real photon $\gamma^*$ and the Pb nucleus, the $\gamma^*$+Pb collision is highly asymmetric and the violation of boost invariance is expected to be even greater than that in p+Pb collisions.
The rapidity decorrelation of the collision geometry thus plays an essential role when computing and measuring the magnitudes of anisotropic flow coefficients in $\gamma^*$+Pb and p+Pb collisions.
Second, the small sizes of the collision systems are pushing the limit of the applicability of the causal relativistic viscous hydrodynamic description~\cite{Bemfica:2020xym, Shen:2020mgh, Chiu:2021muk, Plumberg:2021bme}. Resolving these challenges would bring new exciting opportunities to study collectivity in small systems at the upcoming Electron-Ion Collider (EIC), where one has experimental control on the virtuality of the colliding photon and can use it as a dial to change the collision system size.

In the present work, we explore the collective QGP signatures in $\gamma^*$+Pb collisions at the LHC by employing a full (3+1)D dynamical framework with hydrodynamics and hadronic transport~\cite{Shen:2022oyg}. This framework was shown to provide a unified and quantitative description of particle production in proton-proton, proton-nucleus, and nucleus-nucleus collisions across center of mass energies of a few GeV to several TeV~\cite{Shen:2022oyg}.
We calibrate this model framework with the p+Pb measurements at a center of mass energy $\sqrt{s}=5.02\,{\rm TeV}$ and then make predictions for $\gamma^*$+Pb collisions in the UPC Pb+Pb events.

Our study provides the first quantitative predictions of the anisotropic flow hierarchy between $\gamma^*$+Pb and p+Pb collisions from a final-state-dominated theoretical framework. We also present results for the full rapidity dependence of particle yields and predictions for the photon virtuality dependence of mid-rapidity elliptic flow.

\bigskip
%%%% model
\noindent {\it 2. Methodology. }
In photon-nucleus, $\gamma^*$+Pb, collisions, the virtual photon state may be decomposed into a set of vector meson (VM) states, like $\rho$, $\omega$, and $\phi$ in the vector meson dominance picture \cite{Sakurai:1960ju}. Here, we treat the virtual photon as a vector meson with a lifetime longer than the time of interactions in the low virtuality regime, $Q^2 \sim \Lambda^2_{\rm QCD} - 1$ GeV$^2$~\cite{Shi:2020djm}. Under these assumptions, the photon-nucleus interaction in UPCs proceeds as a vector meson-nucleus collision at an energy lower than that of the associated nucleus-nucleus collisions.
Following Ref.~\cite{Baltz:2007kq}, we consider fluctuating kinematics of the photon-nucleus systems in ultra-peripheral A+A collisions. The probability distribution of the center-of-mass collision energy is 
\begin{eqnarray}\label{eq:rootsdist}
    P(\sqrt{s_{\gamma^*N}}) \propto && \frac{1}{\sqrt{s_{\gamma^*N}}} \bigg[ w_R^{AA}K_0(w_R^{AA}) K_1(w_R^{AA}) \\ \nonumber
    && \qquad - {(w_R^{AA})^2\over 2} \big(K_1^2(w_R^{AA})-K_0^2(w_R^{AA})\big) \bigg],
\end{eqnarray}
where $w_R^{AA} = 2 k R_A/\gamma_L$ with $k= s_{\gamma^*N}/(2\sqrt{s_{\rm NN}})$ and $\gamma_L = \sqrt{s_{\rm NN}}/(2 m_N)$. For the Pb nucleus, we use a Woods-Saxon radius of $R_A = 6.62$ fm, and $K_0(w)$ and $K_1(w)$ are the modified Bessel functions.
In Pb+Pb UPCs at $\sqrt{s_{\rm NN}} = 5.02$ TeV, the $\sqrt{s_{\gamma^*N}}$ of the $\gamma^*$+Pb collisions fluctuates from 0 up to $\sim 894$\,GeV~\cite{Baltz:2007kq,Engel:1994vs,ATLAS:2021jhn}.
%This estimate is consistent with the value given by the Monte Carlo event generator DPMJET-III~\cite{Engel:1994vs, ATLAS:2021jhn}.
To make theoretical predictions with fluctuating $\sqrt{s_{\gamma^*N}}$, we employ the Monte-Carlo \Glauber{} initial-state model~\cite{Shen:2017bsr, Shen:2022oyg} to provide dynamical source terms of energy, momentum, and net baryon density for the subsequent relativistic viscous hydrodynamics evolution. This model can quantitatively reproduce particle production in heavy-ion and asymmetric collisions at different collision energies \cite{Shen:2022oyg}.

We treat the virtual photon as a vector meson and sample two ``partonic participants'' inside the vector meson to capture its geometry shape fluctuations. Those ``partonic participants'' are understood as hot spots composed of valence quarks and their associated gluon cloud. Since the constituent parton distribution functions (PDF) for vector mesons are not well constrained by experiments, we parametrize the vector meson constituent PDF as $xv^{\rm VM}=N_v x^{\alpha} (1-x)^{\beta}$, with $\alpha=\beta =2$, and $N_v$ is subject to the constraint $\int^1_0 v^{\rm VM} (x)dx = 1$. This parametrization is consistent with the shape of the $\pi$ meson's valence PDF at low $Q^2$~\cite{Aicher:2010cb}. The two ``partonic participants'' do not carry all the energy and momentum of the vector meson. The remaining energy and momentum carried by sea quarks and soft gluons are attributed to a soft gluon cloud~\cite{Shen:2022oyg}, which is allowed to participate in collisions with the Pb nucleus. We leave the exploration of UPC observables' sensitivity to the vector meson PDF for future work. The transverse positions of the constituent partons are sampled from a 2D Gaussian, $\exp\left[ - \frac{x^2 + y^2}{2} Q^2 \right]$. We use a default $Q^2 = 0.0625$\,GeV$^2$ in our model calculations, which translates to a vector meson size of 0.8 fm, close to the values for $\rho$ mesons in Ref.~\cite{Forshaw:2003ki}. Below, we will explore the sensitivity of elliptic flow on the photon's virtuality in $\gamma^*$+Pb collisions.

The collective expansion of the produced dense nuclear matter and the evolution of the conserved net-baryon current are simulated within a (3+1)D viscous hydrodynamic framework, \MUSIC~\cite{Schenke:2010rr,Schenke:2010nt,Paquet:2015lta,Denicol:2018wdp}. As the QGP droplet expands and transitions to the dilute hadronic phase, the fluid dynamic description is switched to a microscopic hadron transport model on a constant energy density surface at $e_\mathrm{sw} = 0.20$\,GeV/fm$^3$. The hadronic  transport phase is simulated by \UrQMD~\cite{Bass:1998ca,Bleicher:1999xi} for the hadron scatterings and resonance decays. The numerical simulations of this hybrid \GlauberMUSICUrQMD{} model are carried out using the open-source \iEBEMUSIC{} framework \cite{iEBEMUSIC}. All model parameters were fixed in Ref.~\cite{Shen:2022oyg} by calibrating minimum bias p+p collisions. We adjust the specific shear viscosity to $\eta T/(e+P)=0.08$ and the width of strings in the transverse plane $\sigma_x =0.4$ fm to describe the $p_T$-integrated elliptic flow in p+Pb collisions at 5.02 TeV~\cite{ATLAS:2016yzd}. The (rapidity dependent) charged hadron multiplicity distributions studied in \cite{Shen:2022oyg} are not changed by these adjustments.

%%%% RESULTS
\bigskip
\noindent {\it 3. Collectivity in p+Pb and $\gamma^*$+Pb Collisions. }
The \GlauberMUSICUrQMD{} model can successfully predict the rapidity distributions of particle production in relativistic nuclear collisions of different collision energy and nuclear species~\cite{Shen:2022oyg}. Using this predictive power, we compute the collectivity observables in $\gamma^*$+Pb and p+Pb collisions. We perform numerical simulations for these collisions in their center of mass frames. Because of the unequal energies of projectile and target in the laboratory frame, we need to apply a global rapidity shift to all final-state hadrons before comparing the results with experimental measurements. For p+Pb collisions at $\sqrt{s_{\rm NN}}=5.02$ TeV, the rapidity shift is $\Delta y = 0.465$ in the proton-going direction~\cite{ALICE:2013wgn,CMS:2014und,ATLAS:2015hkr}.  For $\gamma^*$+Pb collisions, the rapidity shift is significantly larger, $\Delta y= y_{\rm beam, \sqrt{s_{\gamma^*N}}} - y_{\rm beam, 5.02 TeV}$, and because of the smaller energy of the $\gamma^*$ is in the Pb-going direction, for example $\Delta y = 1.725$ for  $\sqrt{s_{\gamma^*N}}$ =894 GeV.

\begin{figure}[t]
  \includegraphics[scale=0.4]{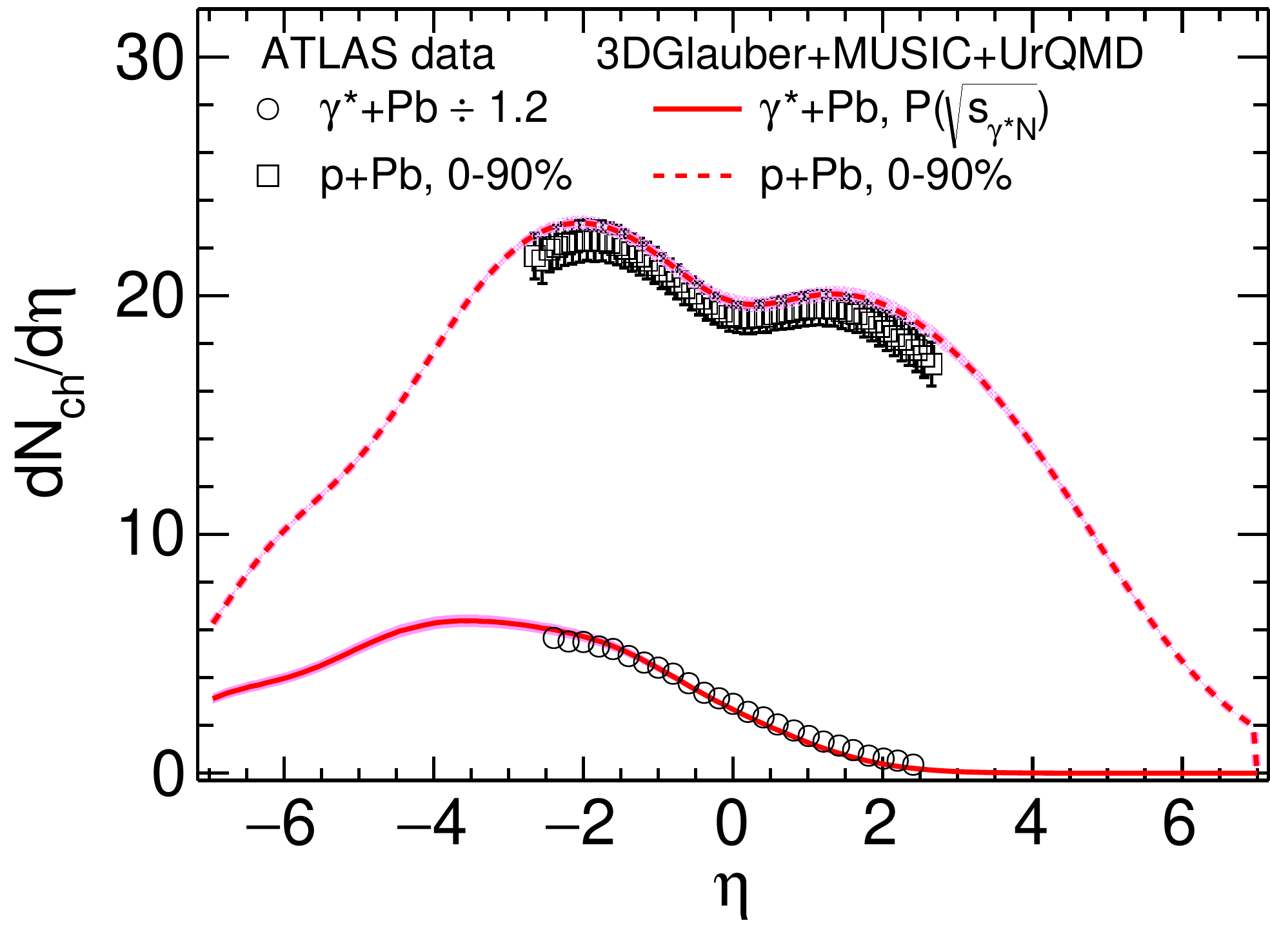}
  \caption{(Color online) The charged hadron pseudorapidity distributions $dN_{\rm ch}/d\eta$ in 0-90\% p+Pb and $\gamma^*$+Pb collisions from the \GlauberMUSICUrQMD{} simulations.  The theoretical calculations are compared with experimental data from the ATLAS Collaboration in the laboratory frame~\cite{ATLAS:2015hkr,ATLAS:2021jhn}. }
  \label{fig:dnchdeta}
\end{figure}
Figure~\ref{fig:dnchdeta} shows our model results for the charged hadron pseudo-rapidity distributions $dN_{\rm ch}/d\eta$ in 0-90\% p+Pb collisions and in $\gamma^*+$Pb collisions for the event class with $N_\mathrm{ch}> 10$ and $0.4 < p_T < 5.0$ GeV in the $\vert \eta \vert < 2.5$ range. Following the ATLAS analysis~\cite{ATLAS:2015hkr}, centrality classes in p+Pb collisions are determined using the total transverse energy measured in the Pb-going direction of the forward calorimeter at $-4.9<\eta<-3.1$. Our model reproduces the shape and magnitude of the p+Pb $dN_{\rm ch}/d\eta$ very well. 

The predicted $dN_{\rm ch}/d\eta$ in $\gamma^*$+Pb collisions shows a strong asymmetry in the $\eta$ direction, which clearly demonstrates the strong violation of longitudinal boost-invariance. %Compared with the ATLAS UPC measurements~\cite{ATLAS:2021jhn}, our result using a fixed $\sqrt{s_{\gamma^*N}}$ = 894 GeV has a slightly flatter slope of $dN_{\rm ch}/d\eta$ than the data.
Implementing fluctuations of $\sqrt{s_{\gamma^*N}}$ according to Eq.\,\eqref{eq:rootsdist} leads to a good description of the shape of the $dN_{\rm ch}/d\eta$ distribution in Pb+Pb UPCs measured by the ATLAS Collaboration. Overall, it is remarkable that a final-state dominated framework can predict the experimental charged hadron rapidity distribution in photo-nuclear events within 10\%. We note that the ATLAS data on $dN_{\rm ch}/d\eta$ in UPCs has been normalized to the value computed by DPMJET-III $\gamma$+Pb at $\eta=0$~\cite{ATLAS:2021jhn}. We divide the ATLAS data of $dN_{\rm ch}/d\eta$ in UPCs by 1.2 to normalize to our result at $\eta=0$.

\begin{figure}[t]
  \includegraphics[scale=0.4]{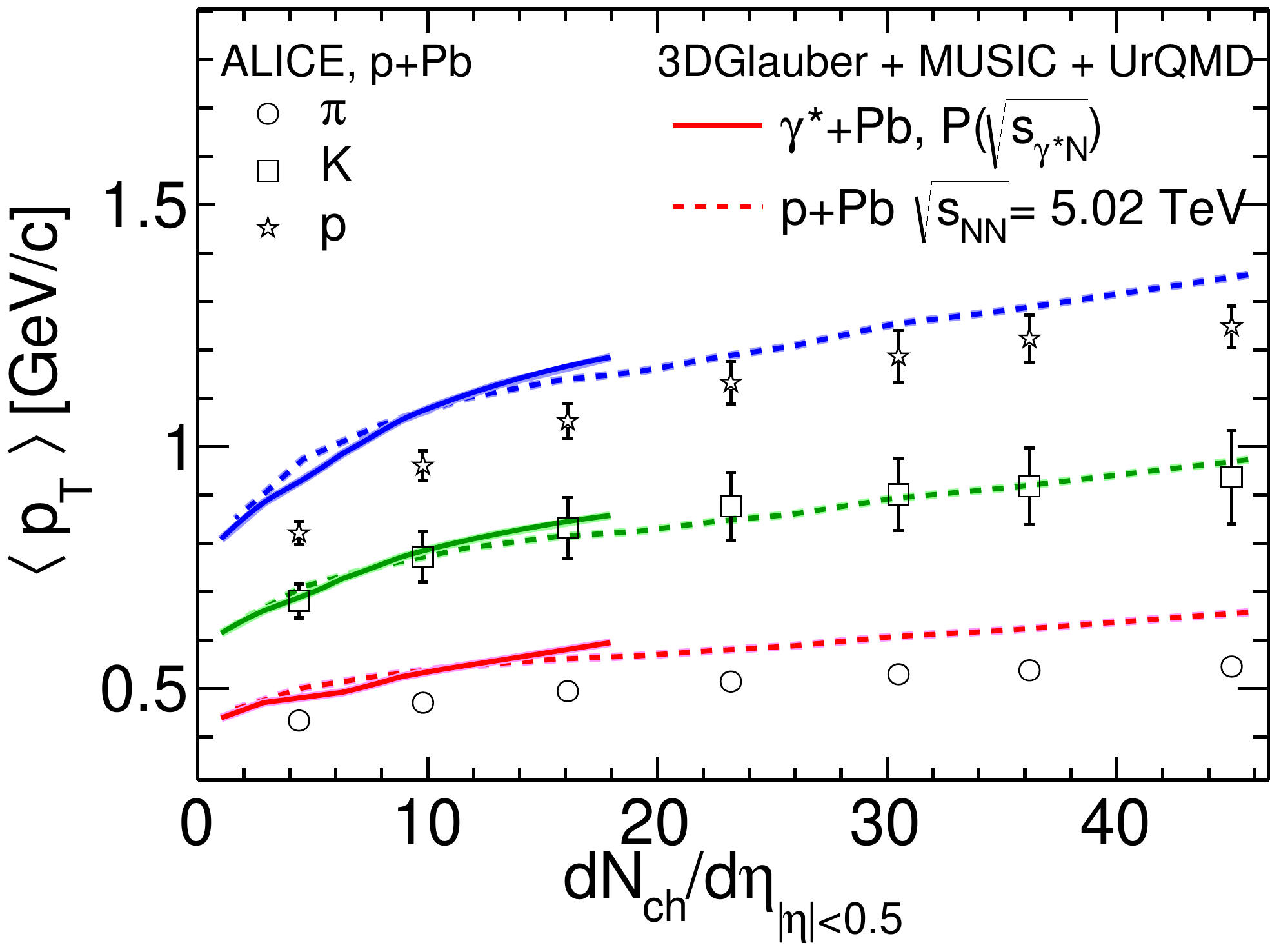}
  \caption{(Color online)  Identified particle mean transverse momenta $\left< p_T\right>$ as functions of charged hadron multiplicity in p+Pb (dashed lines) and $\gamma^*$+Pb (solid lines) collisions from the \GlauberMUSICUrQMD{} framework. The p+Pb results are compared to the experimental data from the ALICE Collaboration~\cite{ALICE:2013wgn}.}
  \label{fig:meanpt}
\end{figure}

The good description of the charged hadron rapidity distributions provides a solid basis for us to quantitatively study flow observables with the same kinematic cuts as done in the experimental analysis. Figure~\ref{fig:meanpt} shows the identified particles' mean transverse momenta $\left< p_T\right>$ as functions of charged hadron multiplicity in p+Pb and $\gamma^*$+Pb collisions. Compared with the ALICE p+Pb measurements~\cite{ALICE:2013wgn}, the \GlauberMUSICUrQMD{} framework reproduces the mass hierarchy of the $\left< p_T\right>$ of pions, kaons, and protons as a result of the system's collective radial expansion. The mean $p_T$ of pions and protons are overestimated by 10\%, which can be improved by including bulk viscous effects in the hydrodynamic evolution. Our model predicts that the identified particles' mean $p_T$ in $\gamma^*$+Pb collisions are very similar to those in p+Pb collisions at the same charged hadron multiplicity. This is a consequence of using the same hot spot size in both systems, which leads to a similar amount of radial flow. Our mean $p_T$ result provides a quantitative prediction and experimental confirmation will be a strong indication that the system produced in $\gamma^*$+Pb collisions experiences strong final-state effects.

\begin{figure}[t]
  \includegraphics[scale=0.4]{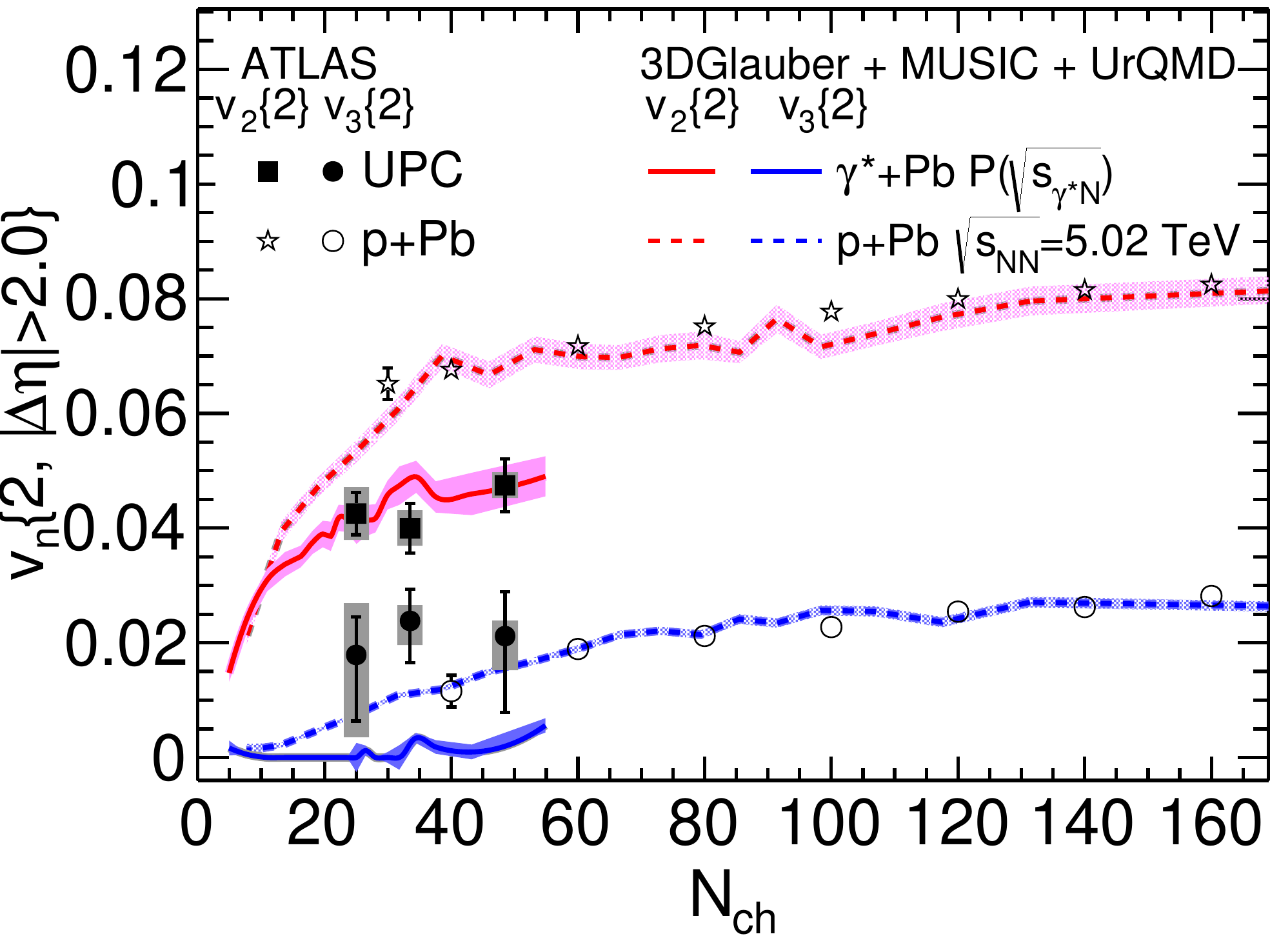}
  \caption{(Color online) Charged hadron anisotropic flow coefficients $v_2\{2\}$ and $v_3\{2\}$ as functions of charged hadron multiplicity $N_{\rm ch}$ in p+Pb (dashed lines) and $\gamma^*$+Pb (solid lines) collisions at LHC energies from the \GlauberMUSICUrQMD{} simulations. To compare with the ATLAS data~\cite{ATLAS:2016yzd,ATLAS:2021jhn}, we compute the $v_n\{2\}$ coefficients using the appropriate kinematic cuts.
  }
  \label{fig:vn2}
\end{figure}

In Figure~\ref{fig:vn2}, we show the multiplicity dependence of the $p_T$-integrated anisotropic flow coefficients $v_2\{2\}$ and $v_3\{2\}$ computed with the Scalar-Product method, which uses two subevents with the kinematic cuts $-2.5<\eta<-1.0$ and $1.0 <\eta<2.5$ and $0.4<p_T<2$\,GeV ($0.5<p_T<5.0$\,GeV) for $\gamma^{*}$+Pb (p+Pb) collisions.
With the specific shear viscosity $\eta T/(e+P) = 0.08$ in hydrodynamic simulations, we fit the ATLAS $v_n$ data for p+Pb collisions \cite{ATLAS:2016yzd}. The \GlauberMUSICUrQMD{} framework can nicely reproduce the multiplicity dependence of the experimentally measured $v_n\{2\}$ in p+Pb collisions. For $N_\mathrm{ch} < 20$, the $v_n\{2\}$ drops because of the decreasing lifetime of the hydrodynamic phase.
Extrapolating from p+Pb to $\gamma^{*}+$Pb collisions, our hydrodynamic calculations reproduce the hierarchy observed for the elliptic flow coefficient $v_2$ in $N_\mathrm{ch} \in [20, 60]$ in the ATLAS data using the template fit method. 

We have checked that the second order eccentricity $\varepsilon_2$ in $\gamma^{*}+$Pb initial states is very close to that in p+Pb systems. However, we find that the longitudinal flow decorrelation is stronger in $\gamma^{*}+$Pb collisions, which results in smaller $v_2\{2\}$ with $\vert \Delta \eta \vert > 2$.  The reasons for the stronger longitudinal flow decorrelation in $\gamma^*$+Pb collisions are (1) the smaller center-of-mass collision energy, which results in a narrower rapidity coverage of particle production, and (2) the larger rapidity shift between the center-of-mass and the lab frames, which increases the decorrelation effects for $\vert \eta \vert < 2.5$ in the lab frame. In other words, the initial transverse geometry is less important than the longitudinal structure in these two small systems.

This result underlines the importance of performing full (3+1)D simulations when quantitatively studying collectivity in small collision systems, and demonstrates that the elliptic flow hierarchy between $\gamma^{*}+$Pb and p+Pb collisions is compatible with a picture where final state effects dominate the generation of momentum anisotropies.

Our model predicts that triangular flow in $\gamma^{*}+$Pb collisions is smaller than that in p+Pb collisions at the same charged hadron multiplicity, again because of the larger longitudinal decorrelation. Consequently, the ordering of $v_3\{2\}$ between $\gamma^{*}+$Pb and p+Pb collisions in our model is opposite to the ATLAS data, which shows a larger $v_3\{2\}$ in $\gamma^{*}+$Pb collisions. The magnitude of $v_3\{2\}$ in $\gamma^{*}+$Pb collisions may be sensitive to vector meson's detailed substructure fluctuations.

\begin{figure}[t]
  \includegraphics[scale=0.4]{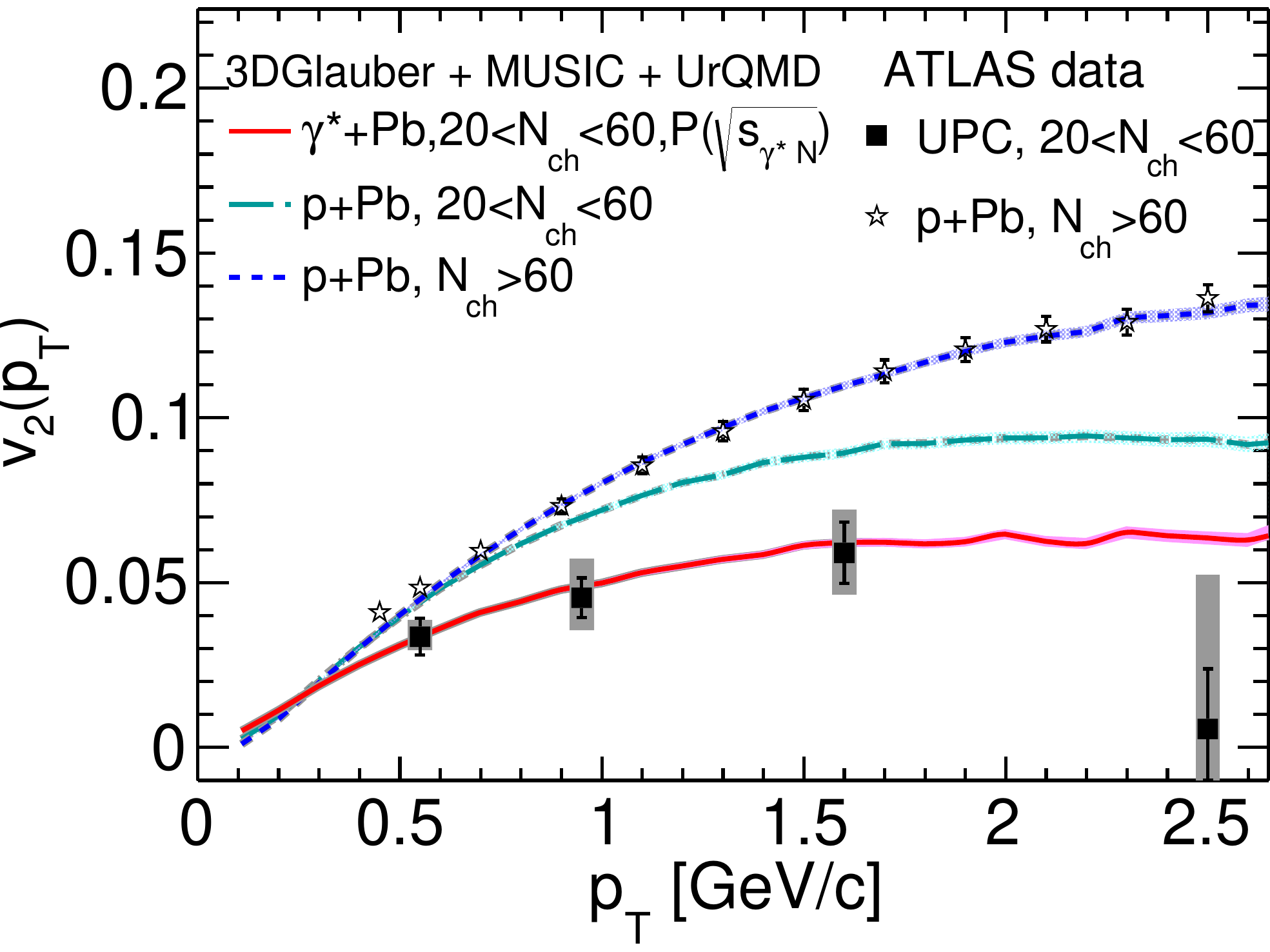}
  \caption{(Color online) The $p_T$-differential elliptic flow coefficient $v_2(p_T)$ of charged hadrons in p+Pb and $\gamma^*$+Pb collisions from the \GlauberMUSICUrQMD{} simulations are compared to ATLAS data~\cite{ATLAS:2016yzd,ATLAS:2021jhn}. The $v_2(p_T)$ are calculated using the Scalar-Product method by imposing $\vert \Delta \eta \vert > 2$ between the particle of interest and the reference charged hadrons in $-2.5<\eta<2.5$ and $0.5<p_T<5.0$ GeV ($0.4<p_T<2.0$ GeV) for p+Pb ($\gamma^*$+Pb) collisions.}
  \label{fig:v2pt}
\end{figure}

Figure~\ref{fig:v2pt} shows our model comparison for the charged hadron $p_T$-differential elliptic flow $v_2(p_T)$ with the ATLAS measurements in $20<N_{\rm ch}<60$ and $N_{\rm ch}> 60$ $\gamma^*$+Pb and p+Pb collisions~\cite{ATLAS:2016yzd,ATLAS:2021jhn}. Our $v_2(p_T)$ result for the $N_{\rm ch}> 60$ events in p+Pb collisions shows excellent agreement with the ALTAS data, marking a good baseline to study the $v_2(p_T)$ in $\gamma^*$+Pb collisions. Comparing this result with the one from the $20<N_{\rm ch}<60$ multiplicity bin of p+Pb collisions, we see a sizable suppression of $v_2(p_T)$ for $p_T > 1$\,GeV in the lower multiplicity bin because of a shorter fireball lifetime.

The $v_2(p_T)$ in $\gamma^*$+Pb collisions in the same $20<N_{\rm ch}<60$ multiplicity bin is 10-15\% smaller than the p+Pb $v_2(p_T)$ across all $p_T$ values because of the larger longitudinal decorrelation with the reference flow angle in $\gamma^*$+Pb collisions. Our model prediction agrees reasonably well with the ATLAS data for $p_T<2.0$ GeV.
The ATLAS UPC $v_2(p_T)$ decreases quickly as $p_T$ increases above 1.6 GeV. This behavior is not seen in our calculations. For $p_T$ above 2 GeV, other physics processes, such as quark recombination, which are not included here, start to be important for anisotropic flow coefficients~\cite{Zhao:2020wcd,Zhao:2021vmu}.

\begin{figure}[t]
  \includegraphics[scale=0.4]{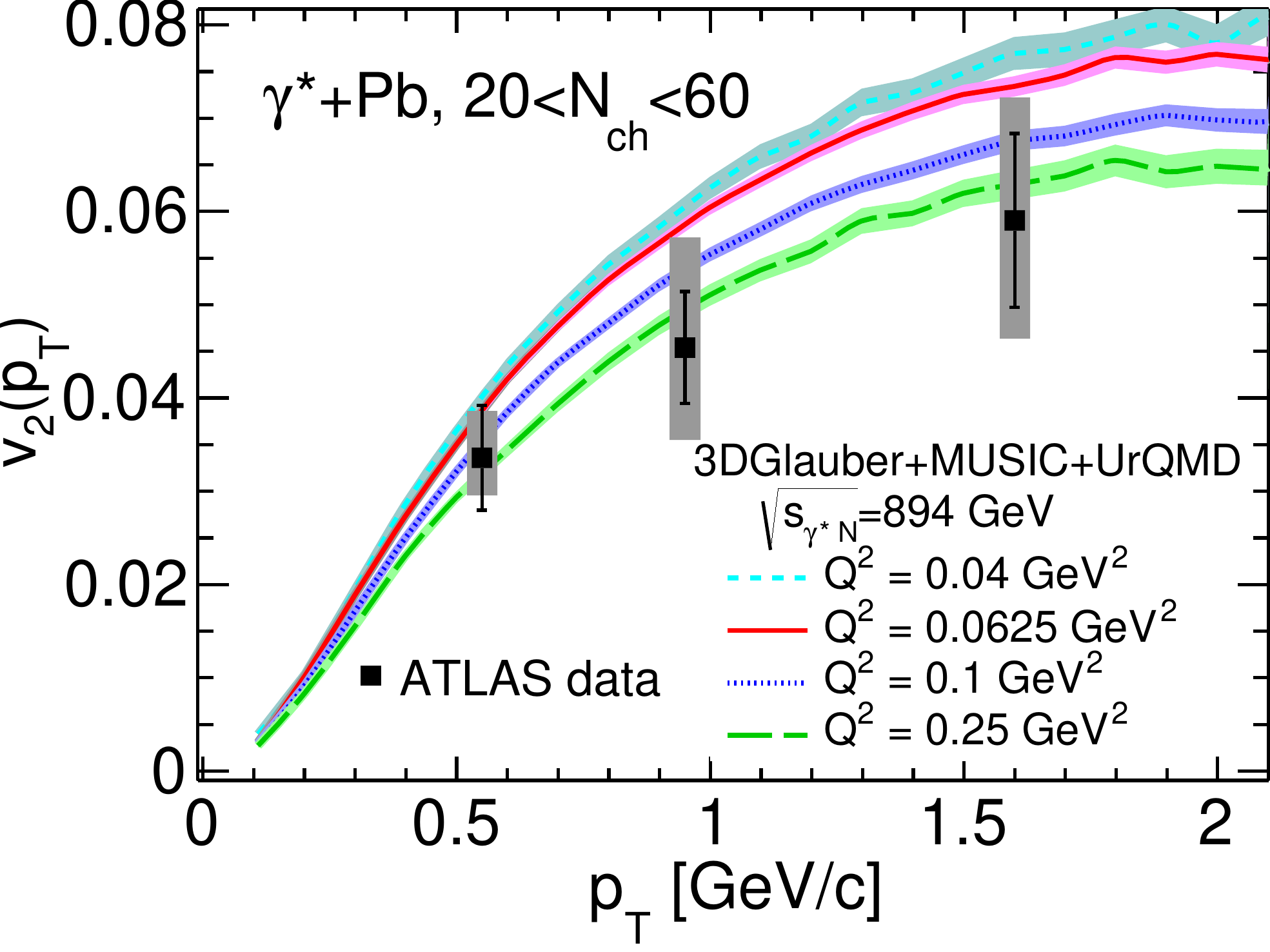}
  \caption{(Color online) The $p_T$-differential elliptic flow coefficient $v_2(p_T)$ of charged hadrons in   $\gamma^*$+Pb collisions from the \GlauberMUSICUrQMD{} simulations with different photon virtualities are compared to the ATLAS  data~\cite{ATLAS:2021jhn}.}
  \label{fig:v2ptdiffBG}
\end{figure}

Finally, we explore the sensitivity of $v_2(p_T)$ to the vector meson transverse size in $\gamma^*$+Pb collisions. Because in the experiment the incoming $\gamma^*$'s virtuality $Q^2$ fluctuates from event to event, the projectile vector meson's average size also fluctuates as it is inversely proportional to $Q^2$. This adds to the geometric fluctuations resulting from the random positions of the two hot spots (at fixed average size). We estimate the uncertainty on the final $v_2(p_T)$ from such $Q^2$ fluctuations by running simulations at different values of $Q^2$.  We perform this exercise at fixed center of mass energy, which requires less statistics. As the $v_2\{2\}$ depends only weakly on the collision energy (see supplemental materials) results for fluctuating energies are expected to be similar.
Figure~\ref{fig:v2ptdiffBG} shows that vector mesons with large virtuality result in smaller elliptic flow coefficients because there is less transverse space for the geometry to fluctuate and the average ellipticities are smaller. Increasing the virtuality from $0.04$\,GeV$^2$ to $0.25$\,GeV$^2$, the $v_2(p_T)$ in $\gamma^*$+Pb decreases monotonically (We remind the reader that te default value is $Q^2 = 0.0625$\,GeV$^2$). The overall relative variation is about 30\%. Future experiments at an Electron-Ion Collider will provide direct access to the photon's virtuality. Therefore, one will be able to systematically test the predictions from the hydrodynamic framework by measuring elliptic anisotropies for different photon virtualities.

\bigskip
%%%% Summary
\noindent {\it 4. Summary. }
In this letter, we have carried out the first dynamical (3+1)D simulations that quantitatively study the  collectivity in p+Pb and ultra-peripheral Pb+Pb collisions at LHC energies within the relativistic hydrodynamic approach. Because these asymmetric collision systems do not have any reasonably wide rapidity window with uniform particle production, it is necessary to go beyond the high-energy Bjorken paradigm and simulate these collisions in full 3D. 

The \GlauberMUSICUrQMD{} model successfully describes the charged hadron pseudo-rapidity distributions in 0-90\% p+Pb and $\gamma^*$+Pb collisions in ultra peripheral Pb+Pb events. This good agreement provides a baseline for studying momentum anisotropies and their correlations in rapidity. 

Although $\gamma^*$+Pb events have slightly larger initial eccentricities compared to those in p+Pb collisions, they also exhibit a stronger longitudinal flow decorrelation, which results in a smaller elliptic momentum anisotropy in $\gamma^*$+Pb collisions in a given multiplicity bin. This result highlights the crucial role of nontrivial longitudinal dynamics in describing the hierarchy of the elliptic flow in p+Pb and  Pb+Pb UPC events.

The simultaneous description of elliptic flow coefficients in p+Pb and $\gamma^*$+Pb collisions in our model supports the hypothesis that strong final state interactions can be the dominant source of collectivity in these small systems. Meanwhile, the fact that the model predicts the opposite trend in $v_3\{2\}$ compared to the experimental data suggests that one needs more careful studies of the vector meson's geometry and the interplay with initial-state correlations.

Our study provides a theoretical framework that can bridge the phenomenological studies of collectivity in relativistic heavy-ion collisions with electron+nucleus collisions at the future Electron-Ion Collider, which will be able to test the picture we have established, namely the dominance of hydrodynamic response to initial spatial geometry in generating azimuthal anisotropies in virtual photon nucleus collisions, with much more precision.

\bigskip
\noindent {\it Acknowledgements.}
We thank Jiangyong Jia, Soumya Mohapatra, and Blair Seidlitz for providing the ATLAS data. W.B.Z. is supported by the National Science Foundation (NSF) under grant numbers ACI-2004571 within the framework of the XSCAPE project of the JETSCAPE collaboration. B.P.S. and C.S. are supported by the U.S. Department of Energy, Office of Science, Office of Nuclear Physics, under DOE Contract No.\,DE-SC0012704 and Award No. DE-SC0021969, respectively. C.S. acknowledges a DOE Office of Science Early Career Award. 
This work is in part supported within the framework of the Beam Energy Scan Theory (BEST) Topical Collaboration and under contract number DE-SC0013460. This research was done using resources provided by the Open Science Grid (OSG) \cite{Pordes:2007zzb, Sfiligoi:2009cct}, which is supported by the National Science Foundation award \#2030508.

\bibliography{references}

\begin{thebibliography}{57}%
\makeatletter
\providecommand \@ifxundefined [1]{%
 \@ifx{#1\undefined}
}%
\providecommand \@ifnum [1]{%
 \ifnum #1\expandafter \@firstoftwo
 \else \expandafter \@secondoftwo
 \fi
}%
\providecommand \@ifx [1]{%
 \ifx #1\expandafter \@firstoftwo
 \else \expandafter \@secondoftwo
 \fi
}%
\providecommand \natexlab [1]{#1}%
\providecommand \enquote  [1]{``#1''}%
\providecommand \bibnamefont  [1]{#1}%
\providecommand \bibfnamefont [1]{#1}%
\providecommand \citenamefont [1]{#1}%
\providecommand \href@noop [0]{\@secondoftwo}%
\providecommand \href [0]{\begingroup \@sanitize@url \@href}%
\providecommand \@href[1]{\@@startlink{#1}\@@href}%
\providecommand \@@href[1]{\endgroup#1\@@endlink}%
\providecommand \@sanitize@url [0]{\catcode `\\12\catcode `\$12\catcode
  `\&12\catcode `\#12\catcode `\^12\catcode `\_12\catcode `\%12\relax}%
\providecommand \@@startlink[1]{}%
\providecommand \@@endlink[0]{}%
\providecommand \url  [0]{\begingroup\@sanitize@url \@url }%
\providecommand \@url [1]{\endgroup\@href {#1}{\urlprefix }}%
\providecommand \urlprefix  [0]{URL }%
\providecommand \Eprint [0]{\href }%
\providecommand \doibase [0]{http://dx.doi.org/}%
\providecommand \selectlanguage [0]{\@gobble}%
\providecommand \bibinfo  [0]{\@secondoftwo}%
\providecommand \bibfield  [0]{\@secondoftwo}%
\providecommand \translation [1]{[#1]}%
\providecommand \BibitemOpen [0]{}%
\providecommand \bibitemStop [0]{}%
\providecommand \bibitemNoStop [0]{.\EOS\space}%
\providecommand \EOS [0]{\spacefactor3000\relax}%
\providecommand \BibitemShut  [1]{\csname bibitem#1\endcsname}%
\let\auto@bib@innerbib\@empty
%</preamble>
\bibitem [{\citenamefont {Heinz}\ and\ \citenamefont
  {Snellings}(2013)}]{Heinz:2013th}%
  \BibitemOpen
  \bibfield  {author} {\bibinfo {author} {\bibfnamefont {U.}~\bibnamefont
  {Heinz}}\ and\ \bibinfo {author} {\bibfnamefont {R.}~\bibnamefont
  {Snellings}},\ }\href {\doibase 10.1146/annurev-nucl-102212-170540}
  {\bibfield  {journal} {\bibinfo  {journal} {Ann. Rev. Nucl. Part. Sci.}\
  }\textbf {\bibinfo {volume} {63}},\ \bibinfo {pages} {123} (\bibinfo {year}
  {2013})},\ \Eprint {http://arxiv.org/abs/1301.2826} {arXiv:1301.2826
  [nucl-th]} \BibitemShut {NoStop}%
\bibitem [{\citenamefont {Gale}\ \emph {et~al.}(2013)\citenamefont {Gale},
  \citenamefont {Jeon},\ and\ \citenamefont {Schenke}}]{Gale:2013da}%
  \BibitemOpen
  \bibfield  {author} {\bibinfo {author} {\bibfnamefont {C.}~\bibnamefont
  {Gale}}, \bibinfo {author} {\bibfnamefont {S.}~\bibnamefont {Jeon}}, \ and\
  \bibinfo {author} {\bibfnamefont {B.}~\bibnamefont {Schenke}},\ }\href
  {\doibase 10.1142/S0217751X13400113} {\bibfield  {journal} {\bibinfo
  {journal} {Int. J. Mod. Phys. A}\ }\textbf {\bibinfo {volume} {28}},\
  \bibinfo {pages} {1340011} (\bibinfo {year} {2013})},\ \Eprint
  {http://arxiv.org/abs/1301.5893} {arXiv:1301.5893 [nucl-th]} \BibitemShut
  {NoStop}%
\bibitem [{\citenamefont {Shen}\ and\ \citenamefont
  {Heinz}(2015)}]{Shen:2015msa}%
  \BibitemOpen
  \bibfield  {author} {\bibinfo {author} {\bibfnamefont {C.}~\bibnamefont
  {Shen}}\ and\ \bibinfo {author} {\bibfnamefont {U.}~\bibnamefont {Heinz}},\
  }\href {\doibase 10.1080/10619127.2015.1006502} {\bibfield  {journal}
  {\bibinfo  {journal} {Nucl. Phys. News}\ }\textbf {\bibinfo {volume} {25}},\
  \bibinfo {pages} {6} (\bibinfo {year} {2015})},\ \Eprint
  {http://arxiv.org/abs/1507.01558} {arXiv:1507.01558 [nucl-th]} \BibitemShut
  {NoStop}%
\bibitem [{\citenamefont {Aidala}\ \emph {et~al.}(2018)\citenamefont {Aidala}
  \emph {et~al.}}]{PHENIX:2017xrm}%
  \BibitemOpen
  \bibfield  {author} {\bibinfo {author} {\bibfnamefont {C.}~\bibnamefont
  {Aidala}} \emph {et~al.} (\bibinfo {collaboration} {PHENIX}),\ }\href
  {\doibase 10.1103/PhysRevLett.120.062302} {\bibfield  {journal} {\bibinfo
  {journal} {Phys. Rev. Lett.}\ }\textbf {\bibinfo {volume} {120}},\ \bibinfo
  {pages} {062302} (\bibinfo {year} {2018})},\ \Eprint
  {http://arxiv.org/abs/1707.06108} {arXiv:1707.06108 [nucl-ex]} \BibitemShut
  {NoStop}%
\bibitem [{\citenamefont {Aidala}\ \emph {et~al.}(2019)\citenamefont {Aidala}
  \emph {et~al.}}]{PHENIX:2018lia}%
  \BibitemOpen
  \bibfield  {author} {\bibinfo {author} {\bibfnamefont {C.}~\bibnamefont
  {Aidala}} \emph {et~al.} (\bibinfo {collaboration} {PHENIX}),\ }\href
  {\doibase 10.1038/s41567-018-0360-0} {\bibfield  {journal} {\bibinfo
  {journal} {Nature Phys.}\ }\textbf {\bibinfo {volume} {15}},\ \bibinfo
  {pages} {214} (\bibinfo {year} {2019})},\ \Eprint
  {http://arxiv.org/abs/1805.02973} {arXiv:1805.02973 [nucl-ex]} \BibitemShut
  {NoStop}%
\bibitem [{\citenamefont {Li}(2012)}]{Li:2012hc}%
  \BibitemOpen
  \bibfield  {author} {\bibinfo {author} {\bibfnamefont {W.}~\bibnamefont
  {Li}},\ }\href {\doibase 10.1142/S0217732312300182} {\bibfield  {journal}
  {\bibinfo  {journal} {Mod. Phys. Lett. A}\ }\textbf {\bibinfo {volume}
  {27}},\ \bibinfo {pages} {1230018} (\bibinfo {year} {2012})},\ \Eprint
  {http://arxiv.org/abs/1206.0148} {arXiv:1206.0148 [nucl-ex]} \BibitemShut
  {NoStop}%
\bibitem [{\citenamefont {Dusling}\ \emph {et~al.}(2016)\citenamefont
  {Dusling}, \citenamefont {Li},\ and\ \citenamefont
  {Schenke}}]{Dusling:2015gta}%
  \BibitemOpen
  \bibfield  {author} {\bibinfo {author} {\bibfnamefont {K.}~\bibnamefont
  {Dusling}}, \bibinfo {author} {\bibfnamefont {W.}~\bibnamefont {Li}}, \ and\
  \bibinfo {author} {\bibfnamefont {B.}~\bibnamefont {Schenke}},\ }\href
  {\doibase 10.1142/S0218301316300022} {\bibfield  {journal} {\bibinfo
  {journal} {Int. J. Mod. Phys. E}\ }\textbf {\bibinfo {volume} {25}},\
  \bibinfo {pages} {1630002} (\bibinfo {year} {2016})},\ \Eprint
  {http://arxiv.org/abs/1509.07939} {arXiv:1509.07939 [nucl-ex]} \BibitemShut
  {NoStop}%
\bibitem [{\citenamefont {Nagle}\ and\ \citenamefont
  {Zajc}(2018)}]{Nagle:2018nvi}%
  \BibitemOpen
  \bibfield  {author} {\bibinfo {author} {\bibfnamefont {J.~L.}\ \bibnamefont
  {Nagle}}\ and\ \bibinfo {author} {\bibfnamefont {W.~A.}\ \bibnamefont
  {Zajc}},\ }\href {\doibase 10.1146/annurev-nucl-101916-123209} {\bibfield
  {journal} {\bibinfo  {journal} {Ann. Rev. Nucl. Part. Sci.}\ }\textbf
  {\bibinfo {volume} {68}},\ \bibinfo {pages} {211} (\bibinfo {year} {2018})},\
  \Eprint {http://arxiv.org/abs/1801.03477} {arXiv:1801.03477 [nucl-ex]}
  \BibitemShut {NoStop}%
\bibitem [{\citenamefont {Nagle}\ \emph {et~al.}(2014)\citenamefont {Nagle},
  \citenamefont {Adare}, \citenamefont {Beckman}, \citenamefont {Koblesky},
  \citenamefont {Orjuela~Koop}, \citenamefont {McGlinchey}, \citenamefont
  {Romatschke}, \citenamefont {Carlson}, \citenamefont {Lynn},\ and\
  \citenamefont {McCumber}}]{Nagle:2013lja}%
  \BibitemOpen
  \bibfield  {author} {\bibinfo {author} {\bibfnamefont {J.~L.}\ \bibnamefont
  {Nagle}}, \bibinfo {author} {\bibfnamefont {A.}~\bibnamefont {Adare}},
  \bibinfo {author} {\bibfnamefont {S.}~\bibnamefont {Beckman}}, \bibinfo
  {author} {\bibfnamefont {T.}~\bibnamefont {Koblesky}}, \bibinfo {author}
  {\bibfnamefont {J.}~\bibnamefont {Orjuela~Koop}}, \bibinfo {author}
  {\bibfnamefont {D.}~\bibnamefont {McGlinchey}}, \bibinfo {author}
  {\bibfnamefont {P.}~\bibnamefont {Romatschke}}, \bibinfo {author}
  {\bibfnamefont {J.}~\bibnamefont {Carlson}}, \bibinfo {author} {\bibfnamefont
  {J.~E.}\ \bibnamefont {Lynn}}, \ and\ \bibinfo {author} {\bibfnamefont
  {M.}~\bibnamefont {McCumber}},\ }\href {\doibase
  10.1103/PhysRevLett.113.112301} {\bibfield  {journal} {\bibinfo  {journal}
  {Phys. Rev. Lett.}\ }\textbf {\bibinfo {volume} {113}},\ \bibinfo {pages}
  {112301} (\bibinfo {year} {2014})},\ \Eprint {http://arxiv.org/abs/1312.4565}
  {arXiv:1312.4565 [nucl-th]} \BibitemShut {NoStop}%
\bibitem [{\citenamefont {Schenke}\ and\ \citenamefont
  {Venugopalan}(2014)}]{Schenke:2014zha}%
  \BibitemOpen
  \bibfield  {author} {\bibinfo {author} {\bibfnamefont {B.}~\bibnamefont
  {Schenke}}\ and\ \bibinfo {author} {\bibfnamefont {R.}~\bibnamefont
  {Venugopalan}},\ }\href {\doibase 10.1103/PhysRevLett.113.102301} {\bibfield
  {journal} {\bibinfo  {journal} {Phys. Rev. Lett.}\ }\textbf {\bibinfo
  {volume} {113}},\ \bibinfo {pages} {102301} (\bibinfo {year} {2014})},\
  \Eprint {http://arxiv.org/abs/1405.3605} {arXiv:1405.3605 [nucl-th]}
  \BibitemShut {NoStop}%
\bibitem [{\citenamefont {Shen}\ \emph {et~al.}(2017)\citenamefont {Shen},
  \citenamefont {Paquet}, \citenamefont {Denicol}, \citenamefont {Jeon},\ and\
  \citenamefont {Gale}}]{Shen:2016zpp}%
  \BibitemOpen
  \bibfield  {author} {\bibinfo {author} {\bibfnamefont {C.}~\bibnamefont
  {Shen}}, \bibinfo {author} {\bibfnamefont {J.-F.}\ \bibnamefont {Paquet}},
  \bibinfo {author} {\bibfnamefont {G.~S.}\ \bibnamefont {Denicol}}, \bibinfo
  {author} {\bibfnamefont {S.}~\bibnamefont {Jeon}}, \ and\ \bibinfo {author}
  {\bibfnamefont {C.}~\bibnamefont {Gale}},\ }\href {\doibase
  10.1103/PhysRevC.95.014906} {\bibfield  {journal} {\bibinfo  {journal} {Phys.
  Rev. C}\ }\textbf {\bibinfo {volume} {95}},\ \bibinfo {pages} {014906}
  (\bibinfo {year} {2017})},\ \Eprint {http://arxiv.org/abs/1609.02590}
  {arXiv:1609.02590 [nucl-th]} \BibitemShut {NoStop}%
\bibitem [{\citenamefont {Weller}\ and\ \citenamefont
  {Romatschke}(2017)}]{Weller:2017tsr}%
  \BibitemOpen
  \bibfield  {author} {\bibinfo {author} {\bibfnamefont {R.~D.}\ \bibnamefont
  {Weller}}\ and\ \bibinfo {author} {\bibfnamefont {P.}~\bibnamefont
  {Romatschke}},\ }\href {\doibase 10.1016/j.physletb.2017.09.077} {\bibfield
  {journal} {\bibinfo  {journal} {Phys. Lett. B}\ }\textbf {\bibinfo {volume}
  {774}},\ \bibinfo {pages} {351} (\bibinfo {year} {2017})},\ \Eprint
  {http://arxiv.org/abs/1701.07145} {arXiv:1701.07145 [nucl-th]} \BibitemShut
  {NoStop}%
\bibitem [{\citenamefont {M\"antysaari}\ \emph {et~al.}(2017)\citenamefont
  {M\"antysaari}, \citenamefont {Schenke}, \citenamefont {Shen},\ and\
  \citenamefont {Tribedy}}]{Mantysaari:2017cni}%
  \BibitemOpen
  \bibfield  {author} {\bibinfo {author} {\bibfnamefont {H.}~\bibnamefont
  {M\"antysaari}}, \bibinfo {author} {\bibfnamefont {B.}~\bibnamefont
  {Schenke}}, \bibinfo {author} {\bibfnamefont {C.}~\bibnamefont {Shen}}, \
  and\ \bibinfo {author} {\bibfnamefont {P.}~\bibnamefont {Tribedy}},\ }\href
  {\doibase 10.1016/j.physletb.2017.07.038} {\bibfield  {journal} {\bibinfo
  {journal} {Phys. Lett. B}\ }\textbf {\bibinfo {volume} {772}},\ \bibinfo
  {pages} {681} (\bibinfo {year} {2017})},\ \Eprint
  {http://arxiv.org/abs/1705.03177} {arXiv:1705.03177 [nucl-th]} \BibitemShut
  {NoStop}%
\bibitem [{\citenamefont {Zhao}\ \emph {et~al.}(2018)\citenamefont {Zhao},
  \citenamefont {Zhou}, \citenamefont {Xu}, \citenamefont {Deng},\ and\
  \citenamefont {Song}}]{Zhao:2017rgg}%
  \BibitemOpen
  \bibfield  {author} {\bibinfo {author} {\bibfnamefont {W.}~\bibnamefont
  {Zhao}}, \bibinfo {author} {\bibfnamefont {Y.}~\bibnamefont {Zhou}}, \bibinfo
  {author} {\bibfnamefont {H.}~\bibnamefont {Xu}}, \bibinfo {author}
  {\bibfnamefont {W.}~\bibnamefont {Deng}}, \ and\ \bibinfo {author}
  {\bibfnamefont {H.}~\bibnamefont {Song}},\ }\href {\doibase
  10.1016/j.physletb.2018.03.022} {\bibfield  {journal} {\bibinfo  {journal}
  {Phys. Lett. B}\ }\textbf {\bibinfo {volume} {780}},\ \bibinfo {pages} {495}
  (\bibinfo {year} {2018})},\ \Eprint {http://arxiv.org/abs/1801.00271}
  {arXiv:1801.00271 [nucl-th]} \BibitemShut {NoStop}%
\bibitem [{\citenamefont {Bierlich}\ and\ \citenamefont
  {Rasmussen}(2019)}]{Bierlich:2019wld}%
  \BibitemOpen
  \bibfield  {author} {\bibinfo {author} {\bibfnamefont {C.}~\bibnamefont
  {Bierlich}}\ and\ \bibinfo {author} {\bibfnamefont {C.~O.}\ \bibnamefont
  {Rasmussen}},\ }\href {\doibase 10.1007/JHEP10(2019)026} {\bibfield
  {journal} {\bibinfo  {journal} {JHEP}\ }\textbf {\bibinfo {volume} {10}},\
  \bibinfo {pages} {026} (\bibinfo {year} {2019})},\ \Eprint
  {http://arxiv.org/abs/1907.12871} {arXiv:1907.12871 [hep-ph]} \BibitemShut
  {NoStop}%
\bibitem [{\citenamefont {Zhao}\ \emph {et~al.}(2020)\citenamefont {Zhao},
  \citenamefont {Ko}, \citenamefont {Liu}, \citenamefont {Qin},\ and\
  \citenamefont {Song}}]{Zhao:2020wcd}%
  \BibitemOpen
  \bibfield  {author} {\bibinfo {author} {\bibfnamefont {W.}~\bibnamefont
  {Zhao}}, \bibinfo {author} {\bibfnamefont {C.~M.}\ \bibnamefont {Ko}},
  \bibinfo {author} {\bibfnamefont {Y.-X.}\ \bibnamefont {Liu}}, \bibinfo
  {author} {\bibfnamefont {G.-Y.}\ \bibnamefont {Qin}}, \ and\ \bibinfo
  {author} {\bibfnamefont {H.}~\bibnamefont {Song}},\ }\href {\doibase
  10.1103/PhysRevLett.125.072301} {\bibfield  {journal} {\bibinfo  {journal}
  {Phys. Rev. Lett.}\ }\textbf {\bibinfo {volume} {125}},\ \bibinfo {pages}
  {072301} (\bibinfo {year} {2020})},\ \Eprint
  {http://arxiv.org/abs/1911.00826} {arXiv:1911.00826 [nucl-th]} \BibitemShut
  {NoStop}%
\bibitem [{\citenamefont {Dusling}\ and\ \citenamefont
  {Venugopalan}(2013)}]{Dusling:2012cg}%
  \BibitemOpen
  \bibfield  {author} {\bibinfo {author} {\bibfnamefont {K.}~\bibnamefont
  {Dusling}}\ and\ \bibinfo {author} {\bibfnamefont {R.}~\bibnamefont
  {Venugopalan}},\ }\href {\doibase 10.1103/PhysRevD.87.051502} {\bibfield
  {journal} {\bibinfo  {journal} {Phys. Rev. D}\ }\textbf {\bibinfo {volume}
  {87}},\ \bibinfo {pages} {051502} (\bibinfo {year} {2013})},\ \Eprint
  {http://arxiv.org/abs/1210.3890} {arXiv:1210.3890 [hep-ph]} \BibitemShut
  {NoStop}%
\bibitem [{\citenamefont {Lappi}\ \emph {et~al.}(2016)\citenamefont {Lappi},
  \citenamefont {Schenke}, \citenamefont {Schlichting},\ and\ \citenamefont
  {Venugopalan}}]{Lappi:2015vta}%
  \BibitemOpen
  \bibfield  {author} {\bibinfo {author} {\bibfnamefont {T.}~\bibnamefont
  {Lappi}}, \bibinfo {author} {\bibfnamefont {B.}~\bibnamefont {Schenke}},
  \bibinfo {author} {\bibfnamefont {S.}~\bibnamefont {Schlichting}}, \ and\
  \bibinfo {author} {\bibfnamefont {R.}~\bibnamefont {Venugopalan}},\ }\href
  {\doibase 10.1007/JHEP01(2016)061} {\bibfield  {journal} {\bibinfo  {journal}
  {JHEP}\ }\textbf {\bibinfo {volume} {01}},\ \bibinfo {pages} {061} (\bibinfo
  {year} {2016})},\ \Eprint {http://arxiv.org/abs/1509.03499} {arXiv:1509.03499
  [hep-ph]} \BibitemShut {NoStop}%
\bibitem [{\citenamefont {Schenke}\ \emph {et~al.}(2016)\citenamefont
  {Schenke}, \citenamefont {Schlichting}, \citenamefont {Tribedy},\ and\
  \citenamefont {Venugopalan}}]{Schenke:2016lrs}%
  \BibitemOpen
  \bibfield  {author} {\bibinfo {author} {\bibfnamefont {B.}~\bibnamefont
  {Schenke}}, \bibinfo {author} {\bibfnamefont {S.}~\bibnamefont
  {Schlichting}}, \bibinfo {author} {\bibfnamefont {P.}~\bibnamefont
  {Tribedy}}, \ and\ \bibinfo {author} {\bibfnamefont {R.}~\bibnamefont
  {Venugopalan}},\ }\href {\doibase 10.1103/PhysRevLett.117.162301} {\bibfield
  {journal} {\bibinfo  {journal} {Phys. Rev. Lett.}\ }\textbf {\bibinfo
  {volume} {117}},\ \bibinfo {pages} {162301} (\bibinfo {year} {2016})},\
  \Eprint {http://arxiv.org/abs/1607.02496} {arXiv:1607.02496 [hep-ph]}
  \BibitemShut {NoStop}%
\bibitem [{\citenamefont {Dusling}\ \emph
  {et~al.}(2018{\natexlab{a}})\citenamefont {Dusling}, \citenamefont {Mace},\
  and\ \citenamefont {Venugopalan}}]{Dusling:2017dqg}%
  \BibitemOpen
  \bibfield  {author} {\bibinfo {author} {\bibfnamefont {K.}~\bibnamefont
  {Dusling}}, \bibinfo {author} {\bibfnamefont {M.}~\bibnamefont {Mace}}, \
  and\ \bibinfo {author} {\bibfnamefont {R.}~\bibnamefont {Venugopalan}},\
  }\href {\doibase 10.1103/PhysRevLett.120.042002} {\bibfield  {journal}
  {\bibinfo  {journal} {Phys. Rev. Lett.}\ }\textbf {\bibinfo {volume} {120}},\
  \bibinfo {pages} {042002} (\bibinfo {year} {2018}{\natexlab{a}})},\ \Eprint
  {http://arxiv.org/abs/1705.00745} {arXiv:1705.00745 [hep-ph]} \BibitemShut
  {NoStop}%
\bibitem [{\citenamefont {Dusling}\ \emph
  {et~al.}(2018{\natexlab{b}})\citenamefont {Dusling}, \citenamefont {Mace},\
  and\ \citenamefont {Venugopalan}}]{Dusling:2017aot}%
  \BibitemOpen
  \bibfield  {author} {\bibinfo {author} {\bibfnamefont {K.}~\bibnamefont
  {Dusling}}, \bibinfo {author} {\bibfnamefont {M.}~\bibnamefont {Mace}}, \
  and\ \bibinfo {author} {\bibfnamefont {R.}~\bibnamefont {Venugopalan}},\
  }\href {\doibase 10.1103/PhysRevD.97.016014} {\bibfield  {journal} {\bibinfo
  {journal} {Phys. Rev. D}\ }\textbf {\bibinfo {volume} {97}},\ \bibinfo
  {pages} {016014} (\bibinfo {year} {2018}{\natexlab{b}})},\ \Eprint
  {http://arxiv.org/abs/1706.06260} {arXiv:1706.06260 [hep-ph]} \BibitemShut
  {NoStop}%
\bibitem [{\citenamefont {Schenke}\ \emph {et~al.}(2020)\citenamefont
  {Schenke}, \citenamefont {Shen},\ and\ \citenamefont
  {Tribedy}}]{Schenke:2019pmk}%
  \BibitemOpen
  \bibfield  {author} {\bibinfo {author} {\bibfnamefont {B.}~\bibnamefont
  {Schenke}}, \bibinfo {author} {\bibfnamefont {C.}~\bibnamefont {Shen}}, \
  and\ \bibinfo {author} {\bibfnamefont {P.}~\bibnamefont {Tribedy}},\ }\href
  {\doibase 10.1016/j.physletb.2020.135322} {\bibfield  {journal} {\bibinfo
  {journal} {Phys. Lett. B}\ }\textbf {\bibinfo {volume} {803}},\ \bibinfo
  {pages} {135322} (\bibinfo {year} {2020})},\ \Eprint
  {http://arxiv.org/abs/1908.06212} {arXiv:1908.06212 [nucl-th]} \BibitemShut
  {NoStop}%
\bibitem [{\citenamefont {Giacalone}\ \emph {et~al.}(2020)\citenamefont
  {Giacalone}, \citenamefont {Schenke},\ and\ \citenamefont
  {Shen}}]{Giacalone:2020byk}%
  \BibitemOpen
  \bibfield  {author} {\bibinfo {author} {\bibfnamefont {G.}~\bibnamefont
  {Giacalone}}, \bibinfo {author} {\bibfnamefont {B.}~\bibnamefont {Schenke}},
  \ and\ \bibinfo {author} {\bibfnamefont {C.}~\bibnamefont {Shen}},\ }\href
  {\doibase 10.1103/PhysRevLett.125.192301} {\bibfield  {journal} {\bibinfo
  {journal} {Phys. Rev. Lett.}\ }\textbf {\bibinfo {volume} {125}},\ \bibinfo
  {pages} {192301} (\bibinfo {year} {2020})},\ \Eprint
  {http://arxiv.org/abs/2006.15721} {arXiv:2006.15721 [nucl-th]} \BibitemShut
  {NoStop}%
\bibitem [{\citenamefont {Schenke}(2021)}]{Schenke:2021mxx}%
  \BibitemOpen
  \bibfield  {author} {\bibinfo {author} {\bibfnamefont {B.}~\bibnamefont
  {Schenke}},\ }\href {\doibase 10.1088/1361-6633/ac14c9} {\bibfield  {journal}
  {\bibinfo  {journal} {Rept. Prog. Phys.}\ }\textbf {\bibinfo {volume} {84}},\
  \bibinfo {pages} {082301} (\bibinfo {year} {2021})},\ \Eprint
  {http://arxiv.org/abs/2102.11189} {arXiv:2102.11189 [nucl-th]} \BibitemShut
  {NoStop}%
\bibitem [{\citenamefont {Aad}\ \emph {et~al.}(2021)\citenamefont {Aad} \emph
  {et~al.}}]{ATLAS:2021jhn}%
  \BibitemOpen
  \bibfield  {author} {\bibinfo {author} {\bibfnamefont {G.}~\bibnamefont
  {Aad}} \emph {et~al.} (\bibinfo {collaboration} {ATLAS}),\ }\href {\doibase
  10.1103/PhysRevC.104.014903} {\bibfield  {journal} {\bibinfo  {journal}
  {Phys. Rev. C}\ }\textbf {\bibinfo {volume} {104}},\ \bibinfo {pages}
  {014903} (\bibinfo {year} {2021})},\ \Eprint
  {http://arxiv.org/abs/2101.10771} {arXiv:2101.10771 [nucl-ex]} \BibitemShut
  {NoStop}%
\bibitem [{\citenamefont {Bertulani}\ \emph {et~al.}(2005)\citenamefont
  {Bertulani}, \citenamefont {Klein},\ and\ \citenamefont
  {Nystrand}}]{Bertulani:2005ru}%
  \BibitemOpen
  \bibfield  {author} {\bibinfo {author} {\bibfnamefont {C.~A.}\ \bibnamefont
  {Bertulani}}, \bibinfo {author} {\bibfnamefont {S.~R.}\ \bibnamefont
  {Klein}}, \ and\ \bibinfo {author} {\bibfnamefont {J.}~\bibnamefont
  {Nystrand}},\ }\href {\doibase 10.1146/annurev.nucl.55.090704.151526}
  {\bibfield  {journal} {\bibinfo  {journal} {Ann. Rev. Nucl. Part. Sci.}\
  }\textbf {\bibinfo {volume} {55}},\ \bibinfo {pages} {271} (\bibinfo {year}
  {2005})},\ \Eprint {http://arxiv.org/abs/nucl-ex/0502005}
  {arXiv:nucl-ex/0502005} \BibitemShut {NoStop}%
\bibitem [{\citenamefont {Baltz}(2008)}]{Baltz:2007kq}%
  \BibitemOpen
  \bibfield  {author} {\bibinfo {author} {\bibfnamefont {A.~J.}\ \bibnamefont
  {Baltz}},\ }\href {\doibase 10.1016/j.physrep.2007.12.001} {\bibfield
  {journal} {\bibinfo  {journal} {Phys. Rept.}\ }\textbf {\bibinfo {volume}
  {458}},\ \bibinfo {pages} {1} (\bibinfo {year} {2008})},\ \Eprint
  {http://arxiv.org/abs/0706.3356} {arXiv:0706.3356 [nucl-ex]} \BibitemShut
  {NoStop}%
\bibitem [{\citenamefont {Bozek}\ \emph {et~al.}(2015)\citenamefont {Bozek},
  \citenamefont {Bzdak},\ and\ \citenamefont {Ma}}]{Bozek:2015swa}%
  \BibitemOpen
  \bibfield  {author} {\bibinfo {author} {\bibfnamefont {P.}~\bibnamefont
  {Bozek}}, \bibinfo {author} {\bibfnamefont {A.}~\bibnamefont {Bzdak}}, \ and\
  \bibinfo {author} {\bibfnamefont {G.-L.}\ \bibnamefont {Ma}},\ }\href
  {\doibase 10.1016/j.physletb.2015.06.007} {\bibfield  {journal} {\bibinfo
  {journal} {Phys. Lett. B}\ }\textbf {\bibinfo {volume} {748}},\ \bibinfo
  {pages} {301} (\bibinfo {year} {2015})},\ \Eprint
  {http://arxiv.org/abs/1503.03655} {arXiv:1503.03655 [hep-ph]} \BibitemShut
  {NoStop}%
\bibitem [{\citenamefont {Ke}\ \emph {et~al.}(2017)\citenamefont {Ke},
  \citenamefont {Moreland}, \citenamefont {Bernhard},\ and\ \citenamefont
  {Bass}}]{Ke:2016jrd}%
  \BibitemOpen
  \bibfield  {author} {\bibinfo {author} {\bibfnamefont {W.}~\bibnamefont
  {Ke}}, \bibinfo {author} {\bibfnamefont {J.~S.}\ \bibnamefont {Moreland}},
  \bibinfo {author} {\bibfnamefont {J.~E.}\ \bibnamefont {Bernhard}}, \ and\
  \bibinfo {author} {\bibfnamefont {S.~A.}\ \bibnamefont {Bass}},\ }\href
  {\doibase 10.1103/PhysRevC.96.044912} {\bibfield  {journal} {\bibinfo
  {journal} {Phys. Rev. C}\ }\textbf {\bibinfo {volume} {96}},\ \bibinfo
  {pages} {044912} (\bibinfo {year} {2017})},\ \Eprint
  {http://arxiv.org/abs/1610.08490} {arXiv:1610.08490 [nucl-th]} \BibitemShut
  {NoStop}%
\bibitem [{\citenamefont {Schenke}\ and\ \citenamefont
  {Schlichting}(2016)}]{Schenke:2016ksl}%
  \BibitemOpen
  \bibfield  {author} {\bibinfo {author} {\bibfnamefont {B.}~\bibnamefont
  {Schenke}}\ and\ \bibinfo {author} {\bibfnamefont {S.}~\bibnamefont
  {Schlichting}},\ }\href {\doibase 10.1103/PhysRevC.94.044907} {\bibfield
  {journal} {\bibinfo  {journal} {Phys. Rev. C}\ }\textbf {\bibinfo {volume}
  {94}},\ \bibinfo {pages} {044907} (\bibinfo {year} {2016})},\ \Eprint
  {http://arxiv.org/abs/1605.07158} {arXiv:1605.07158 [hep-ph]} \BibitemShut
  {NoStop}%
\bibitem [{\citenamefont {Shen}\ and\ \citenamefont
  {Alzhrani}(2020)}]{Shen:2020jwv}%
  \BibitemOpen
  \bibfield  {author} {\bibinfo {author} {\bibfnamefont {C.}~\bibnamefont
  {Shen}}\ and\ \bibinfo {author} {\bibfnamefont {S.}~\bibnamefont
  {Alzhrani}},\ }\href {\doibase 10.1103/PhysRevC.102.014909} {\bibfield
  {journal} {\bibinfo  {journal} {Phys. Rev. C}\ }\textbf {\bibinfo {volume}
  {102}},\ \bibinfo {pages} {014909} (\bibinfo {year} {2020})},\ \Eprint
  {http://arxiv.org/abs/2003.05852} {arXiv:2003.05852 [nucl-th]} \BibitemShut
  {NoStop}%
\bibitem [{\citenamefont {Wu}\ and\ \citenamefont {Qin}(2021)}]{Wu:2021hkv}%
  \BibitemOpen
  \bibfield  {author} {\bibinfo {author} {\bibfnamefont {X.-Y.}\ \bibnamefont
  {Wu}}\ and\ \bibinfo {author} {\bibfnamefont {G.-Y.}\ \bibnamefont {Qin}},\
  }\href@noop {} {\  (\bibinfo {year} {2021})},\ \Eprint
  {http://arxiv.org/abs/2109.03512} {arXiv:2109.03512 [hep-ph]} \BibitemShut
  {NoStop}%
\bibitem [{\citenamefont {Bemfica}\ \emph {et~al.}(2021)\citenamefont
  {Bemfica}, \citenamefont {Disconzi}, \citenamefont {Hoang}, \citenamefont
  {Noronha},\ and\ \citenamefont {Radosz}}]{Bemfica:2020xym}%
  \BibitemOpen
  \bibfield  {author} {\bibinfo {author} {\bibfnamefont {F.~S.}\ \bibnamefont
  {Bemfica}}, \bibinfo {author} {\bibfnamefont {M.~M.}\ \bibnamefont
  {Disconzi}}, \bibinfo {author} {\bibfnamefont {V.}~\bibnamefont {Hoang}},
  \bibinfo {author} {\bibfnamefont {J.}~\bibnamefont {Noronha}}, \ and\
  \bibinfo {author} {\bibfnamefont {M.}~\bibnamefont {Radosz}},\ }\href
  {\doibase 10.1103/PhysRevLett.126.222301} {\bibfield  {journal} {\bibinfo
  {journal} {Phys. Rev. Lett.}\ }\textbf {\bibinfo {volume} {126}},\ \bibinfo
  {pages} {222301} (\bibinfo {year} {2021})},\ \Eprint
  {http://arxiv.org/abs/2005.11632} {arXiv:2005.11632 [hep-th]} \BibitemShut
  {NoStop}%
\bibitem [{\citenamefont {Shen}\ and\ \citenamefont
  {Yan}(2020)}]{Shen:2020mgh}%
  \BibitemOpen
  \bibfield  {author} {\bibinfo {author} {\bibfnamefont {C.}~\bibnamefont
  {Shen}}\ and\ \bibinfo {author} {\bibfnamefont {L.}~\bibnamefont {Yan}},\
  }\href {\doibase 10.1007/s41365-020-00829-z} {\bibfield  {journal} {\bibinfo
  {journal} {Nucl. Sci. Tech.}\ }\textbf {\bibinfo {volume} {31}},\ \bibinfo
  {pages} {122} (\bibinfo {year} {2020})},\ \Eprint
  {http://arxiv.org/abs/2010.12377} {arXiv:2010.12377 [nucl-th]} \BibitemShut
  {NoStop}%
\bibitem [{\citenamefont {Chiu}\ and\ \citenamefont
  {Shen}(2021)}]{Chiu:2021muk}%
  \BibitemOpen
  \bibfield  {author} {\bibinfo {author} {\bibfnamefont {C.}~\bibnamefont
  {Chiu}}\ and\ \bibinfo {author} {\bibfnamefont {C.}~\bibnamefont {Shen}},\
  }\href {\doibase 10.1103/PhysRevC.103.064901} {\bibfield  {journal} {\bibinfo
   {journal} {Phys. Rev. C}\ }\textbf {\bibinfo {volume} {103}},\ \bibinfo
  {pages} {064901} (\bibinfo {year} {2021})},\ \Eprint
  {http://arxiv.org/abs/2103.09848} {arXiv:2103.09848 [nucl-th]} \BibitemShut
  {NoStop}%
\bibitem [{\citenamefont {Plumberg}\ \emph {et~al.}(2021)\citenamefont
  {Plumberg}, \citenamefont {Almaalol}, \citenamefont {Dore}, \citenamefont
  {Noronha},\ and\ \citenamefont {Noronha-Hostler}}]{Plumberg:2021bme}%
  \BibitemOpen
  \bibfield  {author} {\bibinfo {author} {\bibfnamefont {C.}~\bibnamefont
  {Plumberg}}, \bibinfo {author} {\bibfnamefont {D.}~\bibnamefont {Almaalol}},
  \bibinfo {author} {\bibfnamefont {T.}~\bibnamefont {Dore}}, \bibinfo {author}
  {\bibfnamefont {J.}~\bibnamefont {Noronha}}, \ and\ \bibinfo {author}
  {\bibfnamefont {J.}~\bibnamefont {Noronha-Hostler}},\ }\href@noop {} {\
  (\bibinfo {year} {2021})},\ \Eprint {http://arxiv.org/abs/2103.15889}
  {arXiv:2103.15889 [nucl-th]} \BibitemShut {NoStop}%
\bibitem [{\citenamefont {Shen}\ and\ \citenamefont
  {Schenke}(2022)}]{Shen:2022oyg}%
  \BibitemOpen
  \bibfield  {author} {\bibinfo {author} {\bibfnamefont {C.}~\bibnamefont
  {Shen}}\ and\ \bibinfo {author} {\bibfnamefont {B.}~\bibnamefont {Schenke}},\
  }\href@noop {} {\  (\bibinfo {year} {2022})},\ \Eprint
  {http://arxiv.org/abs/2203.04685} {arXiv:2203.04685 [nucl-th]} \BibitemShut
  {NoStop}%
\bibitem [{\citenamefont {Sakurai}(1960)}]{Sakurai:1960ju}%
  \BibitemOpen
  \bibfield  {author} {\bibinfo {author} {\bibfnamefont {J.~J.}\ \bibnamefont
  {Sakurai}},\ }\href {\doibase 10.1016/0003-4916(60)90126-3} {\bibfield
  {journal} {\bibinfo  {journal} {Annals Phys.}\ }\textbf {\bibinfo {volume}
  {11}},\ \bibinfo {pages} {1} (\bibinfo {year} {1960})}\BibitemShut {NoStop}%
\bibitem [{\citenamefont {Shi}\ \emph {et~al.}(2021)\citenamefont {Shi},
  \citenamefont {Wang}, \citenamefont {Wei}, \citenamefont {Xiao},\ and\
  \citenamefont {Zheng}}]{Shi:2020djm}%
  \BibitemOpen
  \bibfield  {author} {\bibinfo {author} {\bibfnamefont {Y.}~\bibnamefont
  {Shi}}, \bibinfo {author} {\bibfnamefont {L.}~\bibnamefont {Wang}}, \bibinfo
  {author} {\bibfnamefont {S.-Y.}\ \bibnamefont {Wei}}, \bibinfo {author}
  {\bibfnamefont {B.-W.}\ \bibnamefont {Xiao}}, \ and\ \bibinfo {author}
  {\bibfnamefont {L.}~\bibnamefont {Zheng}},\ }\href {\doibase
  10.1103/PhysRevD.103.054017} {\bibfield  {journal} {\bibinfo  {journal}
  {Phys. Rev. D}\ }\textbf {\bibinfo {volume} {103}},\ \bibinfo {pages}
  {054017} (\bibinfo {year} {2021})},\ \Eprint
  {http://arxiv.org/abs/2008.03569} {arXiv:2008.03569 [hep-ph]} \BibitemShut
  {NoStop}%
\bibitem [{\citenamefont {Engel}(1995)}]{Engel:1994vs}%
  \BibitemOpen
  \bibfield  {author} {\bibinfo {author} {\bibfnamefont {R.}~\bibnamefont
  {Engel}},\ }\href {\doibase 10.1007/BF01496594} {\bibfield  {journal}
  {\bibinfo  {journal} {Z. Phys. C}\ }\textbf {\bibinfo {volume} {66}},\
  \bibinfo {pages} {203} (\bibinfo {year} {1995})}\BibitemShut {NoStop}%
\bibitem [{\citenamefont {Shen}\ and\ \citenamefont
  {Schenke}(2018)}]{Shen:2017bsr}%
  \BibitemOpen
  \bibfield  {author} {\bibinfo {author} {\bibfnamefont {C.}~\bibnamefont
  {Shen}}\ and\ \bibinfo {author} {\bibfnamefont {B.}~\bibnamefont {Schenke}},\
  }\href {\doibase 10.1103/PhysRevC.97.024907} {\bibfield  {journal} {\bibinfo
  {journal} {Phys. Rev. C}\ }\textbf {\bibinfo {volume} {97}},\ \bibinfo
  {pages} {024907} (\bibinfo {year} {2018})},\ \Eprint
  {http://arxiv.org/abs/1710.00881} {arXiv:1710.00881 [nucl-th]} \BibitemShut
  {NoStop}%
\bibitem [{\citenamefont {Aicher}\ \emph {et~al.}(2010)\citenamefont {Aicher},
  \citenamefont {Schafer},\ and\ \citenamefont {Vogelsang}}]{Aicher:2010cb}%
  \BibitemOpen
  \bibfield  {author} {\bibinfo {author} {\bibfnamefont {M.}~\bibnamefont
  {Aicher}}, \bibinfo {author} {\bibfnamefont {A.}~\bibnamefont {Schafer}}, \
  and\ \bibinfo {author} {\bibfnamefont {W.}~\bibnamefont {Vogelsang}},\ }\href
  {\doibase 10.1103/PhysRevLett.105.252003} {\bibfield  {journal} {\bibinfo
  {journal} {Phys. Rev. Lett.}\ }\textbf {\bibinfo {volume} {105}},\ \bibinfo
  {pages} {252003} (\bibinfo {year} {2010})},\ \Eprint
  {http://arxiv.org/abs/1009.2481} {arXiv:1009.2481 [hep-ph]} \BibitemShut
  {NoStop}%
\bibitem [{\citenamefont {Forshaw}\ \emph {et~al.}(2004)\citenamefont
  {Forshaw}, \citenamefont {Sandapen},\ and\ \citenamefont
  {Shaw}}]{Forshaw:2003ki}%
  \BibitemOpen
  \bibfield  {author} {\bibinfo {author} {\bibfnamefont {J.~R.}\ \bibnamefont
  {Forshaw}}, \bibinfo {author} {\bibfnamefont {R.}~\bibnamefont {Sandapen}}, \
  and\ \bibinfo {author} {\bibfnamefont {G.}~\bibnamefont {Shaw}},\ }\href
  {\doibase 10.1103/PhysRevD.69.094013} {\bibfield  {journal} {\bibinfo
  {journal} {Phys. Rev. D}\ }\textbf {\bibinfo {volume} {69}},\ \bibinfo
  {pages} {094013} (\bibinfo {year} {2004})},\ \Eprint
  {http://arxiv.org/abs/hep-ph/0312172} {arXiv:hep-ph/0312172} \BibitemShut
  {NoStop}%
\bibitem [{\citenamefont {Schenke}\ \emph {et~al.}(2011)\citenamefont
  {Schenke}, \citenamefont {Jeon},\ and\ \citenamefont
  {Gale}}]{Schenke:2010rr}%
  \BibitemOpen
  \bibfield  {author} {\bibinfo {author} {\bibfnamefont {B.}~\bibnamefont
  {Schenke}}, \bibinfo {author} {\bibfnamefont {S.}~\bibnamefont {Jeon}}, \
  and\ \bibinfo {author} {\bibfnamefont {C.}~\bibnamefont {Gale}},\ }\href
  {\doibase 10.1103/PhysRevLett.106.042301} {\bibfield  {journal} {\bibinfo
  {journal} {Phys. Rev. Lett.}\ }\textbf {\bibinfo {volume} {106}},\ \bibinfo
  {pages} {042301} (\bibinfo {year} {2011})},\ \Eprint
  {http://arxiv.org/abs/1009.3244} {arXiv:1009.3244 [hep-ph]} \BibitemShut
  {NoStop}%
\bibitem [{\citenamefont {Schenke}\ \emph {et~al.}(2010)\citenamefont
  {Schenke}, \citenamefont {Jeon},\ and\ \citenamefont
  {Gale}}]{Schenke:2010nt}%
  \BibitemOpen
  \bibfield  {author} {\bibinfo {author} {\bibfnamefont {B.}~\bibnamefont
  {Schenke}}, \bibinfo {author} {\bibfnamefont {S.}~\bibnamefont {Jeon}}, \
  and\ \bibinfo {author} {\bibfnamefont {C.}~\bibnamefont {Gale}},\ }\href
  {\doibase 10.1103/PhysRevC.82.014903} {\bibfield  {journal} {\bibinfo
  {journal} {Phys. Rev. C}\ }\textbf {\bibinfo {volume} {82}},\ \bibinfo
  {pages} {014903} (\bibinfo {year} {2010})},\ \Eprint
  {http://arxiv.org/abs/1004.1408} {arXiv:1004.1408 [hep-ph]} \BibitemShut
  {NoStop}%
\bibitem [{\citenamefont {Paquet}\ \emph {et~al.}(2016)\citenamefont {Paquet},
  \citenamefont {Shen}, \citenamefont {Denicol}, \citenamefont {Luzum},
  \citenamefont {Schenke}, \citenamefont {Jeon},\ and\ \citenamefont
  {Gale}}]{Paquet:2015lta}%
  \BibitemOpen
  \bibfield  {author} {\bibinfo {author} {\bibfnamefont {J.-F.}\ \bibnamefont
  {Paquet}}, \bibinfo {author} {\bibfnamefont {C.}~\bibnamefont {Shen}},
  \bibinfo {author} {\bibfnamefont {G.~S.}\ \bibnamefont {Denicol}}, \bibinfo
  {author} {\bibfnamefont {M.}~\bibnamefont {Luzum}}, \bibinfo {author}
  {\bibfnamefont {B.}~\bibnamefont {Schenke}}, \bibinfo {author} {\bibfnamefont
  {S.}~\bibnamefont {Jeon}}, \ and\ \bibinfo {author} {\bibfnamefont
  {C.}~\bibnamefont {Gale}},\ }\href {\doibase 10.1103/PhysRevC.93.044906}
  {\bibfield  {journal} {\bibinfo  {journal} {Phys. Rev. C}\ }\textbf {\bibinfo
  {volume} {93}},\ \bibinfo {pages} {044906} (\bibinfo {year} {2016})},\
  \Eprint {http://arxiv.org/abs/1509.06738} {arXiv:1509.06738 [hep-ph]}
  \BibitemShut {NoStop}%
\bibitem [{\citenamefont {Denicol}\ \emph {et~al.}(2018)\citenamefont
  {Denicol}, \citenamefont {Gale}, \citenamefont {Jeon}, \citenamefont
  {Monnai}, \citenamefont {Schenke},\ and\ \citenamefont
  {Shen}}]{Denicol:2018wdp}%
  \BibitemOpen
  \bibfield  {author} {\bibinfo {author} {\bibfnamefont {G.~S.}\ \bibnamefont
  {Denicol}}, \bibinfo {author} {\bibfnamefont {C.}~\bibnamefont {Gale}},
  \bibinfo {author} {\bibfnamefont {S.}~\bibnamefont {Jeon}}, \bibinfo {author}
  {\bibfnamefont {A.}~\bibnamefont {Monnai}}, \bibinfo {author} {\bibfnamefont
  {B.}~\bibnamefont {Schenke}}, \ and\ \bibinfo {author} {\bibfnamefont
  {C.}~\bibnamefont {Shen}},\ }\href {\doibase 10.1103/PhysRevC.98.034916}
  {\bibfield  {journal} {\bibinfo  {journal} {Phys. Rev. C}\ }\textbf {\bibinfo
  {volume} {98}},\ \bibinfo {pages} {034916} (\bibinfo {year} {2018})},\
  \Eprint {http://arxiv.org/abs/1804.10557} {arXiv:1804.10557 [nucl-th]}
  \BibitemShut {NoStop}%
\bibitem [{\citenamefont {Bass}\ \emph {et~al.}(1998)\citenamefont {Bass} \emph
  {et~al.}}]{Bass:1998ca}%
  \BibitemOpen
  \bibfield  {author} {\bibinfo {author} {\bibfnamefont {S.~A.}\ \bibnamefont
  {Bass}} \emph {et~al.},\ }\href {\doibase 10.1016/S0146-6410(98)00058-1}
  {\bibfield  {journal} {\bibinfo  {journal} {Prog. Part. Nucl. Phys.}\
  }\textbf {\bibinfo {volume} {41}},\ \bibinfo {pages} {255} (\bibinfo {year}
  {1998})},\ \Eprint {http://arxiv.org/abs/nucl-th/9803035}
  {arXiv:nucl-th/9803035} \BibitemShut {NoStop}%
\bibitem [{\citenamefont {Bleicher}\ \emph {et~al.}(1999)\citenamefont
  {Bleicher} \emph {et~al.}}]{Bleicher:1999xi}%
  \BibitemOpen
  \bibfield  {author} {\bibinfo {author} {\bibfnamefont {M.}~\bibnamefont
  {Bleicher}} \emph {et~al.},\ }\href {\doibase 10.1088/0954-3899/25/9/308}
  {\bibfield  {journal} {\bibinfo  {journal} {J. Phys. G}\ }\textbf {\bibinfo
  {volume} {25}},\ \bibinfo {pages} {1859} (\bibinfo {year} {1999})},\ \Eprint
  {http://arxiv.org/abs/hep-ph/9909407} {arXiv:hep-ph/9909407} \BibitemShut
  {NoStop}%
\bibitem [{iEB()}]{iEBEMUSIC}%
  \BibitemOpen
  \href@noop {} {}\bibinfo {note} {The open-source \textsc{iebe-music}
  overarching framework can be download from
  \url{https://github.com/chunshen1987/iEBE-MUSIC}}\BibitemShut {NoStop}%
\bibitem [{\citenamefont {Aaboud}\ \emph {et~al.}(2017)\citenamefont {Aaboud}
  \emph {et~al.}}]{ATLAS:2016yzd}%
  \BibitemOpen
  \bibfield  {author} {\bibinfo {author} {\bibfnamefont {M.}~\bibnamefont
  {Aaboud}} \emph {et~al.} (\bibinfo {collaboration} {ATLAS}),\ }\href
  {\doibase 10.1103/PhysRevC.96.024908} {\bibfield  {journal} {\bibinfo
  {journal} {Phys. Rev. C}\ }\textbf {\bibinfo {volume} {96}},\ \bibinfo
  {pages} {024908} (\bibinfo {year} {2017})},\ \Eprint
  {http://arxiv.org/abs/1609.06213} {arXiv:1609.06213 [nucl-ex]} \BibitemShut
  {NoStop}%
\bibitem [{\citenamefont {Abelev}\ \emph {et~al.}(2014)\citenamefont {Abelev}
  \emph {et~al.}}]{ALICE:2013wgn}%
  \BibitemOpen
  \bibfield  {author} {\bibinfo {author} {\bibfnamefont {B.~B.}\ \bibnamefont
  {Abelev}} \emph {et~al.} (\bibinfo {collaboration} {ALICE}),\ }\href
  {\doibase 10.1016/j.physletb.2013.11.020} {\bibfield  {journal} {\bibinfo
  {journal} {Phys. Lett. B}\ }\textbf {\bibinfo {volume} {728}},\ \bibinfo
  {pages} {25} (\bibinfo {year} {2014})},\ \Eprint
  {http://arxiv.org/abs/1307.6796} {arXiv:1307.6796 [nucl-ex]} \BibitemShut
  {NoStop}%
\bibitem [{\citenamefont {Khachatryan}\ \emph {et~al.}(2015)\citenamefont
  {Khachatryan} \emph {et~al.}}]{CMS:2014und}%
  \BibitemOpen
  \bibfield  {author} {\bibinfo {author} {\bibfnamefont {V.}~\bibnamefont
  {Khachatryan}} \emph {et~al.} (\bibinfo {collaboration} {CMS}),\ }\href
  {\doibase 10.1016/j.physletb.2015.01.034} {\bibfield  {journal} {\bibinfo
  {journal} {Phys. Lett. B}\ }\textbf {\bibinfo {volume} {742}},\ \bibinfo
  {pages} {200} (\bibinfo {year} {2015})},\ \Eprint
  {http://arxiv.org/abs/1409.3392} {arXiv:1409.3392 [nucl-ex]} \BibitemShut
  {NoStop}%
\bibitem [{\citenamefont {Aad}\ \emph {et~al.}(2016)\citenamefont {Aad} \emph
  {et~al.}}]{ATLAS:2015hkr}%
  \BibitemOpen
  \bibfield  {author} {\bibinfo {author} {\bibfnamefont {G.}~\bibnamefont
  {Aad}} \emph {et~al.} (\bibinfo {collaboration} {ATLAS}),\ }\href {\doibase
  10.1140/epjc/s10052-016-4002-3} {\bibfield  {journal} {\bibinfo  {journal}
  {Eur. Phys. J. C}\ }\textbf {\bibinfo {volume} {76}},\ \bibinfo {pages} {199}
  (\bibinfo {year} {2016})},\ \Eprint {http://arxiv.org/abs/1508.00848}
  {arXiv:1508.00848 [hep-ex]} \BibitemShut {NoStop}%
\bibitem [{\citenamefont {Zhao}\ \emph {et~al.}(2022)\citenamefont {Zhao},
  \citenamefont {Ke}, \citenamefont {Chen}, \citenamefont {Luo},\ and\
  \citenamefont {Wang}}]{Zhao:2021vmu}%
  \BibitemOpen
  \bibfield  {author} {\bibinfo {author} {\bibfnamefont {W.}~\bibnamefont
  {Zhao}}, \bibinfo {author} {\bibfnamefont {W.}~\bibnamefont {Ke}}, \bibinfo
  {author} {\bibfnamefont {W.}~\bibnamefont {Chen}}, \bibinfo {author}
  {\bibfnamefont {T.}~\bibnamefont {Luo}}, \ and\ \bibinfo {author}
  {\bibfnamefont {X.-N.}\ \bibnamefont {Wang}},\ }\href {\doibase
  10.1103/PhysRevLett.128.022302} {\bibfield  {journal} {\bibinfo  {journal}
  {Phys. Rev. Lett.}\ }\textbf {\bibinfo {volume} {128}},\ \bibinfo {pages}
  {022302} (\bibinfo {year} {2022})},\ \Eprint
  {http://arxiv.org/abs/2103.14657} {arXiv:2103.14657 [hep-ph]} \BibitemShut
  {NoStop}%
\bibitem [{\citenamefont {Pordes}\ \emph {et~al.}(2007)\citenamefont {Pordes}
  \emph {et~al.}}]{Pordes:2007zzb}%
  \BibitemOpen
  \bibfield  {author} {\bibinfo {author} {\bibfnamefont {R.}~\bibnamefont
  {Pordes}} \emph {et~al.},\ }\href {\doibase 10.1088/1742-6596/78/1/012057}
  {\bibfield  {journal} {\bibinfo  {journal} {J. Phys. Conf. Ser.}\ }\textbf
  {\bibinfo {volume} {78}},\ \bibinfo {pages} {012057} (\bibinfo {year}
  {2007})}\BibitemShut {NoStop}%
\bibitem [{\citenamefont {Sfiligoi}\ \emph {et~al.}(2009)\citenamefont
  {Sfiligoi}, \citenamefont {Bradley}, \citenamefont {Holzman}, \citenamefont
  {Mhashilkar}, \citenamefont {Padhi},\ and\ \citenamefont
  {Wurthwrin}}]{Sfiligoi:2009cct}%
  \BibitemOpen
  \bibfield  {author} {\bibinfo {author} {\bibfnamefont {I.}~\bibnamefont
  {Sfiligoi}}, \bibinfo {author} {\bibfnamefont {D.~C.}\ \bibnamefont
  {Bradley}}, \bibinfo {author} {\bibfnamefont {B.}~\bibnamefont {Holzman}},
  \bibinfo {author} {\bibfnamefont {P.}~\bibnamefont {Mhashilkar}}, \bibinfo
  {author} {\bibfnamefont {S.}~\bibnamefont {Padhi}}, \ and\ \bibinfo {author}
  {\bibfnamefont {F.}~\bibnamefont {Wurthwrin}},\ }\href {\doibase
  10.1109/CSIE.2009.950} {\bibfield  {journal} {\bibinfo  {journal} {WRI World
  Congress}\ }\textbf {\bibinfo {volume} {2}},\ \bibinfo {pages} {428}
  (\bibinfo {year} {2009})}\BibitemShut {NoStop}%
\end{thebibliography}%

\end{document}